\documentclass [11pt]{article}
\usepackage{amsmath,amsthm,amsfonts,amscd,eucal,latexsym,amssymb,mathrsfs}
\usepackage{hyperref}
\usepackage{epsfig}  
\input xy
\xyoption{all}
 
\def\WF{{\text{WF}}}

\def\beq{\begin{eqnarray}}
\def\eeq{\end{eqnarray}}

\newcommand{\abs}[1]{\left|{#1}\right|}

\def\vsp{\vspace{0.2cm}}

\def\sse #1 {\vsp\ifhmode{\par}\fi\refstepcounter{subsection}
  \noindent {\bf\thesubsection}. {\em #1}.\quad
  \addcontentsline{toc}{subsection}{\protect\numberline{\thesubsection} #1}%
  }

\def\ssb #1 {\vsp\ifhmode{\par}\fi\refstepcounter{subsection}
  \noindent {\bf\thesubsection.} {\bf #1.}\quad
  \addcontentsline{toc}{subsection}{\protect\numberline{\thesubsection} #1}%
  }

\def\ssa #1 {\ifhmode{\par}\fi\refstepcounter{subsection}
  \noindent {\bf\thesubsection.} {\bf #1.}\quad
  \addcontentsline{toc}{subsection}{\protect\numberline{\thesubsection} #1}%
  }

\DeclareMathOperator{\sign}{sign}
\DeclareMathOperator{\supp}{supp}
\DeclareMathOperator{\id}{id}

\def\remark {\vsp\ifhmode{\par}\fi\noindent\noindent {\bf Remark:} 
}

\newtheorem{theorem}{Theorem}[section]
\newtheorem{proposition}{Proposition}[section]

\newtheorem{definition}{Definition}[section]

\begin{document}

\hfill{\sl Desy 10-001, ESI 2203, ZMP-HH/09-33 - January 2010} 
\par 
\bigskip 
\par 
\rm 
 
 
\par 
\bigskip 
\LARGE 
\noindent 
{\bf Local causal structures, Hadamard states and the principle of local covariance in quantum field theory.}
\bigskip 
\par 
\rm 
\normalsize 
 

\large
\noindent 
{\bf Claudio Dappiaggi$^{1,a}$},
{\bf Nicola Pinamonti$^{2,b}$}, {\bf Martin Porrmann$^{3,c}$} \\

\par
\small
\noindent $^1$Erwin Schr\"odinger Institut f\"ur Mathematische Physik,
Boltzmanngasse 9, 
A-1090 Wien, Austria.

\vskip .2cm

\noindent $^2$II. Institut f\"ur Theoretische Physik, Universit\"at Hamburg,
Luruper Chaussee 149, 
D-22761 Hamburg, Germany.\smallskip

\vskip .2cm

\noindent$^3$Quantum Research Group, School of Physics, University of KwaZulu-Natal and National Institute for Theoretical Physics, Private Bag X54001, Durban, 4001, South Africa \bigskip

\noindent $^a$claudio.dappiaggi@esi.ac.at,
$^b$nicola.pinamonti@desy.de,  $^c$martin.porrmann@desy.de\\ 
 \normalsize

\par 
 
\rm\normalsize 
\noindent {\small Version of \today}

\rm\normalsize 
 
 
\par 
\bigskip 

\noindent 
\small 
{\bf Abstract}.
In the framework of the algebraic formulation, we discuss and analyse some new features of the local structure of a real scalar quantum field theory in a strongly causal spacetime. In particular we use the properties of the exponential map to set up a local version of a bulk-to-boundary correspondence. The bulk is a suitable subset of a geodesic neighbourhood of any but fixed point $p$ of the underlying background, while the boundary is a part of the future light cone having $p$ as its own tip. In this regime, we provide a novel notion for the extended $*$-algebra of Wick polynomials on the said cone and, on the one hand, we prove that it contains the information of the bulk counterpart via an injective $*$-homomorphism while, on the other hand, 
we associate to it a distinguished state whose pull-back in the bulk is of Hadamard form. 
The main advantage of this point of view arises if one uses the universal properties of the exponential map and of the light cone in order to show that, for any two given backgrounds $M$ and $M'$ and for any two subsets of geodesic neighbourhoods of two arbitrary points, it is possible to engineer the above procedure such that the boundary extended algebras are related via a restriction homomorphism. This allows for the pull-back of boundary states in both spacetimes and, thus, to set up a machinery which permits the comparison of expectation values of local field observables in $M$ and $M'$.
\normalsize

\vskip .3cm

\bigskip 

\newpage

\tableofcontents

\section{Introduction}
\label{sec:1}

In the framework of quantum field theory over curved backgrounds, we
witnessed a considerable series of leaps forward due to a novel use of
advanced mathematical techniques combined with new physical insights
leading to an improved understanding of the underlying foundations of
the theory.  It is far from our intention to give a recollection of
all of them, but we would like to draw attention at least to some of
them.  On the one hand, in \cite{Brunetti:2003aa}, the principle of
general local covariance was formulated leading to the realisation of
a quantum field theory as a covariant functor between the category of
globally hyperbolic (four-dimensional) Lorentzian manifolds with
isometric embeddings as morphisms and the category of
$C^{\ast}$-algebras with unit-preserving monomorphisms as morphisms and
also to the new interpretation of local fields as natural
transformations from compactly supported smooth function to suitable
operators.  On the other hand, the presence of a nontrivial background
comes with the grievous problem of the \textit{a priori} absence of a
sufficiently large symmetry group to identify a natural ground state
as in Minkowski spacetime where Poincar\'e invariance enables this.

Nonetheless, it is still possible to identify a class of physically
relevant states as those fulfilling the so-called Hadamard condition.
This guarantees that the ultraviolet behaviour of the chosen state
mimics that of the Minkowski vacuum at short distances as well as that
the quantum fluctuations of observables such as the smeared components
of the stress-energy tensor are bounded.  From a practical point of
view, the original characterisation of the Hadamard form was realised
by means of the local structure of the integral kernel of the
two-point function of the selected quasi-free state in a suitably
small neighbourhood of a background point.  Unfortunately, such a
criterion is rather difficult to check in a concrete example and a
real step forward has been achieved in
\cite{Radzikowski:1996aa,Radzikowski:1996ab} in which the connection
between the Hadamard condition and the microlocal properties of the
two-point function is proven and fully characterised.

This result has prompted a series of interesting developments in the
analysis of physically relevant states in a curved background, but we
focus mainly on a few recent advances (\textit{cf.}
\cite{Dappiaggi:2006aa,Dappiaggi:2009aa,Dappiaggi:2009ab}) where it
has been shown that, either in asymptotically flat or in cosmological
spacetimes, it is possible to exploit the conformal structure of the
manifold to identify a preferred null submanifold of codimension one,
the conformal boundary.  On the latter it is possible to coherently
encode the information of the bulk algebra of observables and to
identify a state fulfilling suitable uniqueness properties whose
pull-back in the bulk satisfies the Hadamard condition, being at the
same time invariant under all spacetime isometries.

The main problem in the above construction is the need to find a rigid
and global geometric structure which acts as an auxiliary background
out of which the bulk state is constructed.  Hence the local
applicability of a similar scheme seems rather limited; yet one of the
main goals of the present paper is to show that such a procedure can
indeed be set up at a local level and for all spacetimes of physical
interest.  In particular, this statement is established on the basis
of a careful use of some rather well-known geometrical objects.

To be more precise, the point of view taken is the following: if one
considers an arbitrary but fixed point $p$ in a strongly causal
four-dimensional spacetime, it is always possible to single out a
geodesic neighbourhood where the exponential map is a local
diffeomorphism.  Within this set we can also always select a second
point $q$ such that the double cone $\mathscr{D} \equiv \mathscr{D} (
p , q ) \doteq I^{+} ( p ) \cap I^{-} ( q )$ is a globally hyperbolic
spacetime.  This line of reasoning has a twofold advantage: on the one
hand, one can single out a local natural null submanifold of
codimension one, $\mathscr{C}^{+}_{p}$, as the portion of $J^{+} ( p
)$ contained in the closure of $\mathscr{D}$, while, on the other
hand, we are free to repeat the very same construction for a second
point $p^{\prime}$ with associated double cone $\mathscr{D}^{\prime}$
in another spacetime $M^{\prime}$.  Since the exponential map is
invertible and the tangent spaces $T_{p} ( M )$ and $T_{p^{\prime}} (
M^{\prime} )$ are isomorphic, it turns out that it is possible to
engineer all the geometric data in such a way that the two boundaries
$\mathscr{C}^{+}_{p}$ and $\mathscr{C}^{+}_{p^{\prime}}$ can be
related by a suitable restriction map, the only freedom being the
choice of a frame at $p$ and at $p^{\prime}$.

These two advantages can be used to draw some important conclusions on
the structure of local quantum field theories.  More precisely, we
shall focus on a real scalar field theory in $\mathscr{D}$, with
generic mass $m$ and with generic curvature coupling $\xi$.  The
associated quantum observables are described by the Borchers-Uhlmann
algebra or rather by the extended algebra of fields.  In particular,
we shall show that it is possible to construct a scalar field theory
also on $\mathscr{C}_{p}$ and, as a novel result, that also, in the
boundary, there exists a natural notion of extended algebra which is
made precise here.  Apart from the check of mathematical consistency
of our definition, we reinforce our proposal by showing that there
exists an injective $^{\ast}$-homomorphism $\Pi$ between the bulk and
the boundary counterparts.  The relevance of this result is emphasised
by the identification of a natural state on the boundary, whose pull-back in $\mathscr{D}$
via $\Pi$ turns out still to be invariant under a change of the frame
(hence, physically speaking, it is the same for all inertial observers
at $p$) and to be of Hadamard form.  This result provides a potential
candidate for a local vacuum in the large class of backgrounds to be
considered here.

Yet we still have not made profitable use of the second advantage
outlined before.  As a matter of fact, we can now consider two
arbitrary strongly causal spacetimes $M$ and $M^{\prime}$ as well as
two points therein so that the relevant portions of the two
boundaries, $\mathscr{C}_{p}$ and $\mathscr{C}_{p'}$, say, associated
with the double cones, can be related by a suitable restriction map.
The construction of the boundary field theory shows that such a map
becomes an injective homomorphism between the boundary extended
algebras, hence allowing for the construction of a local Hadamard
state in two different backgrounds starting from the same building
block on the boundary.

The results presented in the present work have some antecedents in the
concept of relative Cauchy evolution developed in
\cite{Brunetti:2003aa}.  Knowing a theory (and its corresponding
Hadamard function) in the neighbourhood of a Cauchy surface, such a
method permits one to reconstruct the theory in the neighbourhood of
any other Cauchy surface of the same spacetime.  The deformation
arguments, see \cite{Fulling:1981aa}, play a crucial role in obtaining
the Hadamard property for the deformed state in particular.  Another
related key result is also the one presented in \cite{Verch:1994aa}
about the local quasi-equivalence of quasi-free Hadamard states.  In
the present paper, using the null cones as hypersurfaces on which to
encode the quantum information, we succeed in giving an extended
algebra of observables without knowing the state in a neighbourhood of
such a surface.  Hence, based on the new method presented here, it is
possible to determine quantum states out of their form on null
surfaces alone and thus in a spacetime-independent way.

We are now in a position to have a reference state with respect to
which we can compare the expectation values of the same field
observables in two different spacetimes.  In particular, if one of
these is (a portion of) Minkowski spacetime, it is obvious that the
result of the comparison will be related to the geometric data of the
second background which can now be assessed with a crystal clear
procedure.  Furthermore, we shall show that this method admits an
interpretation within the language of category theory, so that it
becomes manifest that our proposal is not in contrast with the
principle of general local covariance, but can actually be seen as a
generalisation.  As a matter of fact, it reduces to the latter
whenever isometric embeddings are involved, in which case the fields
recover their interpretation as natural transformations as in
\cite{Brunetti:2003aa}, \textit{i.e.}, they transform in a covariant
manner under local isometries.

To reinforce the above procedure we also provide an explicit example
of this ``comparison'' strategy considering a massless real scalar
field minimally coupled to scalar curvature both in Minkowski and in a
Friedman-Robertson-Walker spacetime with flat spatial sections.  We
demonstrate how the difference of the expectation values of the
regularised squared scalar fields in these two spacetimes can be
expanded into a power series of a suitable local coordinate system
(null-advanced) yielding, at first order, a contribution dependent on
the structure of the so-called scale factor of the curved background.

Since we have already extensively discussed the plan of action, we
only briefly sketch the synopsis of the paper.  In Section~\ref{sec:2}
we shall analyse all the geometric structure needed.  Although most of
the material, devoted to the construction of frames and of the
exponential map, is rather well-known in the literature, we try
nonetheless to recollect it here to provide guidance through the
construction of the main geometric objects required, the boundary in
particular.  In Section~\ref{sec:3} we shall tackle the problem of
constructing a quantum scalar field theory on a null cone; in
particular, in Subsections~\ref{ssec:3.1} and \ref{ssec:3.2} we
discuss the structure of the bulk and boundary algebras therein while,
in Subsection~\ref{ssec:3.3}, we identify the distinguished boundary
state.  The novel construction of the extended algebra on the boundary
is presented in Subsection~\ref{ssec:3.4} and all these results are
connected to the bulk counterpart in Subsection~\ref{ssec:3.5}.
Eventually, in Section~\ref{sec:4}, we discuss, by means of the
language of categories, the scheme which leads to the possibility to
compare field theories on different spacetimes.  The concrete example
mentioned above is in Subsection~\ref{ssec:4.2}.  Section~\ref{sec:5}
summarises the paper and sets out a few conclusions as well as
possible future investigations.

\section{Frames and Cones}
\label{sec:2}

As outlined in the introduction, the keyword of this paper is
``comparison,'' \textit{i.e.} our ultimate goal will be to correlate
quantum field theories in different backgrounds both at the level of
algebras and of states and, moreover, to try in the process also to
extract information on the local geometry.  To this avail one needs a
crystal clear control both of the underlying background and of its
properties.  Therefore, we cannot consider arbitrary manifolds, but
need to focus only on those which are of physical relevance insofar as
they can carry a full-fledged quantum field theory.

If we keep in mind this perspective, we shall henceforth call
\emph{spacetime} a four-dimensional, Hausdorff, connected smooth
manifold $M$ endowed with a Lorentzian metric whose signature is
$(-,+,+,+)$.  Then, consequently, $M$ is also second countable and
paracompact \cite{Geroch:1968aa,Geroch:1970aa}.  Customarily one also
requires that $M$ be globally hyperbolic (see for example
\cite{BarGinouxPfaffle_Wave-Equations:2007} or \cite{Brunetti:2003aa})
in order to have a well-defined Cauchy problem for the equations of
motion ruling the dynamics of standard free field theories.

The next natural step is the identification of further local geometric
structures which could serve as a useful tool in the comparison of two
different field theories on two different spacetimes, $M$ and
$M^{\prime}$.  It is known that, for a real scalar field theory, it
suffices to require that the two spacetimes are either isometrically
embedded into each other or conformally related
\cite{Brunetti:2003aa,Pinamonti:2009aa}.  The drawback of this
approach is that only few pairs $M$ and $M^{\prime}$ fulfil such
criteria and potentially interesting cases, such as when $M$ coincides
with Minkowski spacetime and $M^{\prime}$ with de Sitter, are
excluded.

A natural alternative would be to consider pairs of spacetimes $M$ and
$M^{\prime}$ related by a global diffeomorphism, but, unfortunately,
these maps do no preserve the geometric structures at the heart of the
quantum or even of the classical field theory.  A typical example of
such a problem arises in connection with the equations of motion of a
dynamical system whenever these are constructed out of the spacetime
metric.  The action of a generic diffeomorphism preserves their form
only in special cases, \textit{viz.} when they are related to
isometries.  Hence we would return to the original scenario.

Apart from these remarks we should also keep in mind the idea, briefly
sketched in the introduction, to exploit a bulk-to-boundary
reconstruction procedure along the lines of
\cite{Dappiaggi:2006aa,Dappiaggi:2009aa}.  At a global level, this
requires the existence of a conformal boundary structure, a feature
shared only by a certain class of manifolds.  Since we want to
consider a scenario as general as possible, a viable alternative is to
focus only on the \emph{local} structures of the underlying
spacetimes.  In the remainder of this section, we show how to
substantiate this heuristic idea if one carefully uses certain
properties of the exponential map.

\subsection{Frames and the Exponential Map}
\label{ssec:2.1}

The aim of this subsection is to introduce the basic geometric tools
to be used.  Most of the concepts are certainly well-known in the
literature and the reader might refer either to
\cite{Husemoller_Fibre-Bundles:1994} for a full-fledged analysis of
those related to bundles and their properties or to
\cite{kobayashi/nomizu:1963,ONeill_Semi-Riemannian:1983} for a
discussion focused on the differential geometric aspects.
Nonetheless, it is worthwhile to recapitulate part of them since they
will play a pivotal role in this paper and we can, at the same time,
fix the notation.

Consider an arbitrary four-dimensional differentiable manifold $M$.
To any point $p \in M$, we can associate
\begin{itemize}
\item a \emph{linear frame} $F_{p}$ of the tangent space,
  \textit{i.e.}, a non-singular linear mapping $e : \mathbb{R}^{4} \to
  T_{p} ( M )$, or, equivalently, an assignment of an ordered basis
  $e_{1}$, \dots, $ e_{4}$ of $T_{p} ( M )$.
\end{itemize}
It is straightforward to infer that the set of all such linear frames
$FM$ at an arbitrary but fixed $p \in M$ naturally comes with a right
and free action of the group $GL(4,\mathbb{R})$ which is tantamount to
the possible changes of basis in $\mathbb{R}^{4}$, \textit{i.e.}, $( A
, e ) \mapsto eA$ where $eA$ denotes the ordered basis $A^{i}_{j}
e_{i}$ for all $A \in GL(4,\mathbb{R})$.  Thus $FM$ can be endowed
with the following additional structure:
\begin{itemize}
\item Given a four-dimensional differentiable manifold $M$, a
  \emph{frame bundle} is the principal bundle $\widetilde{FM} = F [
  GL(4,\mathbb{R}) , \pi^{\prime} , M ]$ built from the disjoint union
  $\bigsqcup_{p} \widetilde{F_{p} M}$, where $\widetilde{F_{p} M}$ is
  identified with the typical fibre $GL(4,\mathbb{R})$ and
  $\pi^{\prime} : \widetilde{FM} \to M$ is the projection map.
  Furthermore, the tangent bundle $TM$ can be constructed as the
  associated bundle $TM = \widetilde{FM} \times_{GL(4,\mathbb{R})}
  \mathbb{R}^{4}$.
\end{itemize}

We emphasise the well-known fact that the structure introduced last
guarantees that the typical fibre of the tangent bundle at any point
$p$ is $\mathbb{R}^{4}$ regardless of the chosen manifold, a fact we
shall use in the forthcoming discussion.  Following
\cite{kobayashi/nomizu:1963,ONeill_Semi-Riemannian:1983}, recall that
\begin{itemize}
\item for any $p \in M$, if $D_{p}$ is the set of all vectors $v$ in
  $T_{p} ( M )$ such that the geodesic $\gamma_{v} : [ 0 , 1 ] \to M$
  admits $v$ as tangent vector in $0$, then the exponential map at $p$
  is $\exp_{p} : D_{p} \to M$ with $\exp_{p} ( v ) = \gamma_{v} ( 1
  )$;
\item for any point $p \in M$ there always exists a neighbourhood
  $\widetilde{\mathcal{O}}$ of the $0$-vector in $T_{p} ( M )$ such
  that the exponential map is a diffeomorphism onto an open subset
  $\mathcal{O} \subset M$.  Furthermore, whenever
  $\widetilde{\mathcal{O}}$ is star-shaped, $\mathcal{O}$ is called a
  \emph{normal neighbourhood}, and the inverse map therein will be
  denoted $\exp_{p}^{-1} : \mathcal{O} \to \widetilde{\mathcal{O}}$.
\end{itemize}

Although the existence of open sets where the exponential map is a
diffeomorphism suggests a way to compare local quantum field theories
on different manifolds, we also need to single out a preferred
structure of codimension $1$, since we wish to implement a
bulk-to-boundary procedure.  To this avail all the manifolds are
henceforth endowed with a smooth Lorentzian metric, which entails the
following additional features:
\begin{itemize}
\item Since a linear frame at a point $p \in M$ can be seen as the
  assignment of an ordered basis of $\mathbb{R}^{4}$, one can endow
  this latter vector space with the standard Minkowski metric $\eta$,
  which, by construction, is invariant under the Lorentz group
  $SO(3,1)$.  In this case the frame bundle becomes $FM = F [ SO(3,1)
  , \pi^{\prime} , M ]$ which is also referred to as the \emph{bundle
    of orthonormal frames over $M$}.  Furthermore, if the spacetime is
  oriented and time-oriented, we can further reduce the group to
  $SO_{0}(3,1)$, the component of $SO(3,1)$ connected to the identity.
\item Every point in a Lorentzian manifold admits a normal
  neighbourhood (see Proposition~7 and also Definition~5 in Chapter~5
  of \cite{ONeill_Semi-Riemannian:1983}).
\item There is always a choice of coordinates, called \emph{normal
    coordinates}, such that, in these coordinates, the pull-back of
  the metric $g$ under the inverse of the exponential map equals
  $\eta$ (the Minkowski metric in standard coordinates) on the inverse
  image of the point $p$.
\item Since we shall ultimately need to single out a sort of preferred
  codimension $1$ structure, it is rather important that, in a
  Lorentzian manifold, the so-called Gauss lemma holds true (Lemma 1
  in Chapter~5 of \cite{ONeill_Semi-Riemannian:1983}).  In particular,
  this entails that, given any $p \in M$, if we consider the null cone
  $\widetilde{C} \subset T_{p} ( M )$ having $p$ as its own tip, then
  the subset $\widetilde{C} \cap \widetilde{\mathcal{O}}$ is mapped
  into a local null cone in $\mathcal{O} \subset M$ which consists of
  initial segments of all null geodesics starting at $p$.
\end{itemize}

We are now in a position to outline the building blocks of our
geometric construction.  Let us consider two spacetimes $( M , g )$
and $( M^{\prime} , g^{\prime})$ and two generic points $p \in M$ and
$p^{\prime} \in M^{\prime}$, together with their normal neighbourhoods
$\mathcal{O}_{p}$ and $\mathcal{O}_{p^{\prime}}$.  If we equip each
tangent space with an orthonormal basis via a frame, $e :
\mathbb{R}^{4} \to T_{p} ( M )$ and $e^{\prime} : \mathbb{R}^{4} \to
T_{p^{\prime}} ( M^{\prime} )$, we are also free to introduce a map
$i_{e,e^{\prime}} : T_{p} ( M ) \to T_{p^{\prime}} ( M^{\prime} )$
which is constructed simply by identifying the elements of the two
ordered bases.

The strategy is now to exploit the fact that the exponential map is a
diffeomorphism (hence invertible) in a geodesic neighbourhood to
introduce a map $\imath_{e,e^{\prime}} : \mathcal{O}_{p} \to
\mathcal{O}_{p^{\prime}}$ such that
\begin{equation}
  \label{eq:morphisms}
  \imath_{e,e^{\prime}} \doteq \exp_{p^{\prime}} \circ \medspace
  i_{e,e^{\prime}} \circ \exp_{p}^{-1} \text{.} 
\end{equation}
It is important to stress a few further aspects of this last
definition:
\begin{itemize}
\item The map $\imath_{e,e^{\prime}}$ is well defined only when
  $\exp_{p^{\prime}}^{-1}$ can be inverted on the image of
  $i_{e,e^{\prime}} \circ \exp_{p}^{-1}$, that is when
  $i_{e,e^{\prime}} \circ \exp_{p}^{-1} ( \mathcal{O}_{p} ) \subset
  \widetilde{\mathcal{O}}_{p^{\prime}}$.  Therefore, for the sake of
  notational simplicity, when we write $\imath_{e,e^{\prime}}$ it is
  always assumed that such a requirement is satisfied.  Furthermore,
  for every point $p$ we can always consider a sufficiently smaller
  subset of $\mathcal{O}_{p}$, retaining all its properties, where the
  above inclusion holds true.
\item The map $\imath_{e,e^{\prime}}$, which maps a sufficiently small
  $\mathcal{O}$ to $\mathcal{O^{\prime}}$, is not unique, in the sense
  that it depends on the chosen orthonormal frames $e$ and
  $e^{\prime}$.  We have always the freedom to act with an element of
  the structure group of the fibre (be it $SO(3,1)$ or $SO_{0}(3,1)$
  depending on the scenario considered) which maps an orthonormal
  basis into a second one, and this either on $T_{p} ( M )$ or
  $T_{p^{\prime}} ( M^{\prime})$.  Such arbitrariness cannot be lifted
  and, for this reason, we have explicitly indicated the two frames in
  the mapping $\imath_{e,e^{\prime}}$.
\item Despite the freedom mentioned above, the map
  $\imath_{e,e^{\prime}}$ is invariant under the action of a single
  element of the structure group of the fibre on both $e$ and
  $e^{\prime}$, \textit{i.e.}, there exists an equivalence relation:
  We say that
  \begin{equation}
    \label{eq:equivrel}
    \imath_{e,e^{\prime}} \sim \imath_{\tilde{e},\tilde{e}^{\prime}}
  \end{equation}
  if and only if there exists an element $\Lambda \in SO_{0}(3,1)$
  such that $\tilde{e} = \Lambda e$ and $\tilde{e}^{\prime} = \Lambda
  e^{\prime}$.  This equivalence relation shall actually play a
  relevant role in the discussion of Section~\ref{sec:4}.
\end{itemize}

As a related point, notice that, if the spacetime $M$ is isometrically
embedded into $M^{\prime}$, a scenario close to the hypotheses in
\cite{Brunetti:2003aa}, each isometry $\phi : M \to M^{\prime}$
induces an isomorphism between the orthonormal frame bundles $FM$ and
$FM^{\prime}$ since the metric structure is preserved.  In this case
every local character of the manifold $M$ is preserved under $\phi$
(see for example Chapter~3 of \cite{ONeill_Semi-Riemannian:1983}) and,
hence, one can consider a sufficiently small subset of the normal
neighbourhood of any $p \in M$ as well as of $\phi ( p ) \in
M^{\prime}$ so that our construction yields the following commutative
diagram,
\begin{equation*}
  \begin{CD}
    \mathcal{O}_{p} @> {\exp_{p}^{-1}} >> T_{p} ( M ) \\
    @V{\phi}VV        @VV{i_{e,({{\phi}_{\ast}} \circ \thinspace e)}}V \\
    \mathcal{O}_{\phi(p)} @ << {\exp_{{\phi(p)}}} <  T_{\phi(p)} (
    M^{\prime} )
  \end{CD}
  \qquad \raisebox{-2\baselineskip}{\text{.}}
\end{equation*}
Notice that the presence of $\phi_{\ast} \circ e$ in place of a
generic $e^{\prime}$ can be justified as follows: If we call
$(~\thinspace,~)_{p}$ the inner product between vectors in $T_{p} ( M
)$, then for any $v$, $w \in T_{p} ( M )$, one has $( v , w )_{p} = (
\phi_{\ast} ( v ) , \phi_{\ast} ( w ) )_{\phi(p)}$, which, upon
introduction of a local frame $e : \mathbb{R}^{4} \to T_{p} ( M )$,
yields $( v , w )_{p} = ( e ( v_{i} ) , e ( w_{i} ) )_{p} = (
\phi_{\ast} \circ e ( v_{i} ) , \phi_{\ast} \circ e ( w_{i} )
)_{\phi(p)}$ where $v_{i}$, $w_{i} \in \mathbb{R}^{4}$.  Moreover, if
a generic $e^{\prime}$ is used in place of $\phi_{\ast} \circ e$ there
is no guarantee that the previous diagram commutes.  A counterexample
can actually be constructed considering two isometrically related
spacetimes, which are not rotationally invariant and taking for
$e^{\prime}$, $\phi_{\ast} \circ e$ rotated by some generic angle.

\subsection{Double Cones and Their Past Boundary}
\label{ssec:2.2}

The analysis of the previous subsection is a first step towards the
setup of a full-fledged procedure which allows for the local
comparison of quantum field theories on different spacetimes.  We shall
now single out a preferred submanifold of codimension $1$ on which to
apply a bulk-to-boundary reconstruction.

To this avail we have to ensure in the first place that one can
consistently assign to the background $M$ a well-defined quantum field
theory.  Since we are only interested in local quantities, the usual
hypothesis of global hyperbolicity of the spacetime can be moderately
relaxed and, henceforth, we shall assume $M$ to be \emph{strongly
  causal} \cite{beem/ehrlich/easley:1996}, \textit{i.e.}, for every
point $p \in M$, there exists an arbitrarily small convex, causally
convex neighbourhood $\mathcal{O}^{\prime}_p$, which means that no
non-spacelike curve intersects $\mathcal{O}^{\prime}_{p}$ in a
disconnected set.  In other words, $\mathcal{O}^{\prime}_{p}$ itself
is globally hyperbolic.

From a physical point of view, this simply forces us to require that,
ultimately, the theory coincides with the usual quantisation procedure
on each of these subsets, while, from a geometrical perspective, the
discussion of the preceding section still holds true since we are
entitled to select $\mathcal{O}^{\prime}_{p} \subseteq
\mathcal{O}_{p}$, the normal neighbourhood of $p$, in such a way that
the exponential map is a local diffeomorphism also on
$\mathcal{O}^{\prime}_{p}$.  Furthermore, a rather useful class of
sets is constructed out of the so-called \emph{double cones},
\begin{equation*}
  \mathscr{D} ( p^{\prime} , q ) = I^{+} ( p^{\prime} ) \cap I^{-} ( q
  ) \subset M \text{,}
\end{equation*}
where $I^{\pm}$ stand, respectively, for the chronological future and
past while $q \in \mathcal{O}^{\prime}_{p^{\prime}}$.  Notice that
both $p^{\prime}$ and $q$ can be arbitrary but, for our construction,
we shall always suppose that at least one of them coincides with $p$,
henceforth $p^{\prime} \equiv p$.  It is also interesting that
$\mathscr{D} ( p , q )$ is an open and still globally hyperbolic
subset of $\mathcal{O}^{\prime}_{p}$.  In the forthcoming discussion
the boundary of this region will also be relevant and we point out
that the closure $\overline{\mathscr{D} ( p , q )}$ is a compact set
(see for example Chapter 8 in \cite{Wald_General-Relativity:1984})
which coincides with $J^{+} ( p ) \cap J^{-} ( q )$.  Furthermore, it
is also important to recall both that the set of (the closures of)
double cones can be used as a base of the topology of
$\mathcal{O}^{\prime}_{p}$ and that, under the previous assumptions,
we can also freely consider the image of $\overline{\mathscr{D} ( p ,
  q )}$ under the inverse exponential map $\exp^{-1}_{p}$, denoted by
$U ( p , q )$.  The reader should bear in mind that $U ( p , q )$ is
not necessarily the closure of a double cone in $T_{p} ( M ) \sim
\mathbb{R}^{4}$ with respect to the flat metric since only (portions
of) cones in $T_{p} ( M )$, having $p$ as their tip, are mapped in
(portions of) those in $\mathcal{O}_{p}$ and vice versa.

Nonetheless, this construction allows for the identification of the
main geometrical structure needed, since the very existence of
$\mathscr{D} ( p , q )$ and the properties of this set as well as of
$J^{+} ( p )$ under the exponential map suggest to consider
$\mathscr{C}^{+}_{p} \doteq \partial J^{+} ( p ) \cap
\overline{\mathscr{D} ( p , q )}$ as the natural boundary on which to
encode data from a field theory in the bulk.  The bulk here means
$\mathscr{D} ( p , q )$ which is a genuine globally hyperbolic
submanifold of $M$ on which a full-fledged quantum field theory can
indeed be defined.

From a geometrical point of view, a few interesting intrinsic
properties of $\mathscr{C}^{+}_{p}$ can readily be inferred, namely,
to start with, $\mathscr{C}^{+}_{p}$ is generated by future directed
null geodesics in particular originating from $p$.  Notice that the
latter are not complete since the set we are interested in is
constrained to $\overline{\mathscr{D} ( p , q )} \subset
\mathcal{O}^{\prime}_{p}$ and, therefore, its image under
$\exp_{p}^{-1}$ in $T_{p} ( M )$ identifies a portion of a
future directed null cone $C^{+}$ constructed with respect to the flat
metric $\eta$, where this portion is topologically equivalent to $I
\times \mathbb{S}^{2}$, $I \subseteq \mathbb{R}$.  Yet all these
properties are universal, thus they do not depend on the choice of a
specific frame $e$ at $p$.  This is not the case for the form of the
image of $\mathscr{C}^{+}_{p}$ in $C^{+}$ under $\exp_{p}^{-1}$ or the
pull-back of the metric in normal coordinates under $\exp_{p}^{\ast}$.
These clearly depend upon the coordinate system considered
(individuated by $e$) and, hence, the possible choices of $e$ and of
coordinates on $\mathscr{C}^{+}_{p}$ deserve a more detailed
discussion.

If one starts from the observation that the double cones of interest
all lie in a normal neighbourhood, a first natural guess is to select
the standard normal coordinates constructed out of the frame $e$.  In
this setting the metric can be expanded as
\begin{equation*}
  g_{\mu \nu} ( q ) = \eta_{\mu \nu} - \frac{1}{3} R_{\mu \alpha \nu
    \beta} ( p ) \sigma^{\alpha} ( q , p ) \sigma^{\beta} ( q , p ) +
  O ( 3 ) \text{,}
\end{equation*}
where $\sigma ( q , p )$ is the so-called Synge's world function,
\textit{i.e.} half of the square of the geodesic distance between $p$
and $q$ (see Section~2.1 of \cite{Poisson:2004ab}).  Here
$\sigma^{\alpha}$ and $\sigma^{\beta}$ denote the covariant
derivatives of $\sigma$ performed at $p$, whereas $O ( 3 )$ is a
shortcut to stress that the metric is approximated up to cubic
quantities in the normal coordinates.

Unfortunately, both the coordinate system and the expansion are not
well suited to be used in the analysis of the geometry of the null
cone $\mathscr{C}^{+}_{p}$, since one would like to have a local chart
where it is manifest that $\mathscr{C}^{+}_{p}$ is a null
hypersurface.  Furthermore, for later purposes, we also need to
discuss some properties of the metric in the vicinity of
$\mathscr{C}_{p}^{+}$ as a whole and not only in a neighbourhood of
$p$.  To this end, it is useful to employ the so called \emph{retarded
  coordinates} as introduced in \cite{Poisson:2004aa,Preston:2006aa}.
We also refer to the review \cite{Poisson:2004ab}, which has the
advantage to clearly discuss the explicit relation between these new
coordinates and the normal ones (or, also, the Fermi-Walker ones).

Let us briefly recall the construction of these retarded
coordinates. Consider a timelike geodesic $\tilde{\gamma}$ through $p$
with unit tangent vector $u$.  In this setting one can define a
coordinate $r$ as the field
\begin{equation*}
  r ( q ) \doteq - \sigma_{\alpha} ( q , p^{\prime} ) \thinspace
  u^{\alpha} ( p^{\prime} ) \text{,} 
\end{equation*}
where $q \in \mathcal{O}$ and $p^{\prime} \in \tilde{\gamma}$ are
connected by a light-like geodesic originating from $p^{\prime}$ and
pointing towards the future.  With $\sigma_{\alpha} ( q , p^{\prime} )$ we mean the covariant derivative 
at $p$ of the geodesic distance. The
net advantage of $r$ is that, on $\mathscr{C}_{p}^{+}$, it can be read
as an affine parameter of the null geodesics emanating from $p$.  In
other words, once an orthonormal frame $e$ is chosen in such a way
that $e^{0} ( p^{\prime} ) = u$, the scalar field $r$ on
$C^{+}_{p^{\prime}}$ is unambiguously fixed.

We can now define the full retarded coordinates as $( u , r ,x^A )$,
where $u$ labels the family of forward null cones with tips lying on
$\tilde{\gamma}$ (see equation (154) and the preceding discussion in
\cite{Poisson:2004ab}), and
\begin{equation*}
  \mathscr{C}^{+}_{p} = \left\{ p^{\prime} \in \mathscr{D} ( p , q )
    \medspace | \medspace u ( p^{\prime} ) = 0 \right\} \text{,}
\end{equation*}
while $x^A$ are local coordinates on $\mathbb{S}^{2}$.  Notice that
one could alternatively switch to the more common local chart $(
\theta , \varphi )$ of $\mathbb{S}^{2}$ at $p^{\prime}$ and we shall
do so whenever needed.

Moreover, in this coordinate system, the most generic form of the
metric reads \cite{Choquet-Bruhat:2009aa}
\begin{equation}
  \label{eq:metriccone}
  ds^{2} = - \alpha \thinspace du^{2} + 2 \upsilon_{A} du \thinspace
  dx^{A} - 2 e^{2\beta} du \thinspace dr + g^{\prime}_{A B} dx^{A}
  dx^{B} \text{,} 
\end{equation}
where $\alpha$, $\upsilon_{A}$, $\beta$ and $g^\prime_{A B}$ are smooth
functions depending on the coordinates.  Notice that here $r \in ( 0 ,
\infty )$ while $u$ ranges over an open set $I \subseteq \mathbb{R}$
which contains $0$.  Moreover, the $x$-coordinates on the sphere give
rise to a volume element with respect to \eqref{eq:metriccone} of the
form
\begin{equation}
  \label{eq:spherevolumeelement}
  \sqrt{\abs{g^{\prime}_{A B}}} \medspace dx^{A} \wedge dx^{B} =
  \sqrt{\abs{g_{A B}}} \medspace \abs{\sin \theta} \medspace d\theta
  \wedge d\varphi \text{,}
\end{equation}
where the symbol $\abs{~\cdot~}$ under the square root is kept to
recall that we are actually referring to the determinant of the
matrices involved.  Notice also that, depending on the chosen
coordinates $x^{A}$ on $\mathbb{S}^{2}$, the switch to $ ( \theta ,
\varphi )$ yields a harmless additional contribution to the metric
coefficients; this justifies the two symbols $g^{\prime}_{A B}$ and
$g_{A B}$, although, henceforth, we shall mostly stick to the last
one.

It is also remarkable that, whenever $R_{\mu \nu} \dot{\gamma}^{\mu}
\dot{\gamma}^{\nu} = 0$, $\gamma$ a generator of
$\mathscr{C}^{+}_{p}$, one can prove that, on $\mathscr{C}^{+}_{p}$,
\eqref{eq:metriccone} simplifies (see formula (2.36) in
\cite{Choquet-Bruhat:2009aa}) to
\begin{equation}
  \label{eq:metriccone2}
  ds^{2} = - \alpha \thinspace du^{2} - 2 \thinspace du \thinspace dr
  + r^{2} ( d\theta^{2} + \sin^{2} \theta \thinspace d\varphi^{2} )
  \text{,}
\end{equation}
where the standard coordinates $( \theta , \varphi )$ on the
$2$-sphere are used in place of $x^{A}$.  Apart from being much
simpler, this form is more closely related to the standard Bondi one
which is canonically used in the implementation of bulk-to-boundary
techniques as devised in \cite{Dappiaggi:2006aa,Dappiaggi:2009aa} for
a large class of asymptotically flat and of cosmological spacetimes.
Unfortunately, contrary to these papers, here the scenario is much
more complicated and, furthermore, the cone does not seem to display
any particular symmetry group to be exploited, such as for example the
BMS in \cite{Dappiaggi:2006aa}.  Yet, the situation is not as
desperate as one might think, since, ultimately, for our purposes it
will only be relevant that the metric at $p$ becomes the Minkowski
one, our coordinates being constructed out of an orthonormal frame at
$p$.  In particular, this means that, at $p$, $\sqrt{| g_{A B} |}$
will become proportional to $r$, which, in this special scenario, can
be seen both as the affine null parameter introduced above, or,
equivalently, as the standard radial coordinate in Minkowski spacetime
constructed out of the orthonormal frame in $T_{p} ( M ) \sim
\mathbb{R}^{4}$.

Before concluding this section, we briefly compare
\eqref{eq:metriccone2} with the corresponding expression in Minkowski
spacetime, where the flat metric can be written as
\begin{equation}
  \label{eq:flatcone}
  ds^{2} = - dU^{2} + 2 \thinspace dU \thinspace dr + r^{2} (
  d\theta^{2} + \sin^{2} \theta \thinspace  d\varphi^{2} ) \text{,}
\end{equation}
$U \doteq t + r$ denoting the light coordinate constructed out of the
time and spherical coordinates.  The cone with tip at $0$ is
characterised by $U = 0$ and, also in this case, $r$ is an affine
parameter along the null geodesics emanating from $0$.  It is
important to stress that the pull-back of \eqref{eq:metriccone} under
$\exp_{p}^{\ast}$ tends to \eqref{eq:flatcone} when approaching the
point $\exp_{p} ( 0 ) = p$.

Finally, we comment on the behaviour of $\mathscr{D} ( p , q )$ under
\eqref{eq:morphisms}.  As mentioned before, $\exp^{-1}_{p}$ does not
map a double cone in $M$ into one in $T_{p} ( M )$, but, nonetheless,
we can still adapt the choice of $q$ in such a way that $\imath_{e,e'}
( \mathscr{D} ( p , q ) )$ is properly contained in a sufficiently
large double cone $\mathscr{D} ( p , q^{\prime}) \subset
\mathcal{O}^{\prime}_p$.

\section{Algebras of Observables on the Bulk and on the Boundary}
\label{sec:3}

In the previous section, we focused on the introduction and analysis
of the main geometric tools needed.  In particular, we recall once
more that the main geometrical objects are the double cones
$\mathscr{D} ( p , q )$ which are globally hyperbolic spacetimes in
their own right.  Since, in the forthcoming discussion, neither $p$
nor $q$ will play a distinguished role, we shall omit them, hence
using $\mathscr{D}$ in place of $\mathscr{D} ( p , q )$.  More
importantly, we are now entitled to introduce a well-defined classical
field theory and, for the sake of simplicity, we shall henceforth only
deal with a free real scalar field with generic mass $m$ and generic
curvature coupling $\xi$.

Let us recollect some standard properties of such a physical system
along the lines, \textit{e.g.}, of
\cite{Wald_Quantum-Field-Theory:1994}.  Consider $\varphi :
\mathscr{D} \to \mathbb{R}$ which fulfils the following equation of
motion,
\begin{equation}
  \label{eq:eqm}
  P \varphi \doteq \left( \Box_{g} + \xi R + m^{2} \right) \varphi = 0
  \text{,} \quad m^{2} > 0 \text{~and~} \xi \in \mathbb{R} \text{,}
\end{equation}
where $\Box_{g} = - \nabla^{\mu} \nabla_{\mu}$ is the d'Alembert wave
operator constructed out of the metric $g$ while $R$ is the scalar
curvature.  Since this is a second-order hyperbolic partial
differential equation, each solution with smooth and compactly
supported initial data on a Cauchy surface can be constructed as the
image of the following map,
\begin{equation}
  \label{eq:causalp}
  \Delta : C^{\infty}_{0} ( \mathscr{D} ) \to C^{\infty} ( \mathscr{D}
  ) \text{,}
\end{equation}
where $\Delta$ is the \emph{causal propagator} defined as the
difference of the advanced and the retarded fundamental
solutions.  Furthermore, each $\varphi_{f} \doteq \Delta ( f )$ satisfies the
following support property,
\begin{equation*}
  \supp ( \varphi_{f} ) \subseteq J^{+} ( \supp ( f ) ) \cup J^{-} (
  \supp ( f ) ) \text{,}
\end{equation*}
and, if $\mathfrak{S} ( \mathscr{D} )$ denotes the set of solutions of
\eqref{eq:causalp} with smooth compactly supported initial data on any
Cauchy surface $\Sigma$ of $\mathscr{D}$, then this turns out to be a
symplectic space when endowed with the weakly non-degenerate
symplectic form,
\begin{equation}
  \label{eq:symplbulk}
  \sigma ( \varphi_{f} , \varphi_{h} ) = \int_\Sigma d\mu ( \Sigma )
  \left( \varphi_{f} \nabla_{n} \varphi_{h} - \varphi_{h} \nabla_{n}
    \varphi_{f} \right) = \int_\mathscr{D} d\mu ( \mathscr{D} ) ( f
  \thinspace \Delta  h ) \text{,} \quad \forall f \text{,\thinspace} h
  \in C^{\infty}_{0} ( \mathscr{D} ) \text{.}
\end{equation}
Here the integral is independent of the choice of Cauchy surface
$\Sigma$, as can be noticed from the last equation, while $d\mu (
\Sigma )$, $d\mu ( \mathscr{D} )$ and $n$ are, respectively, the
metric-induced measures and the vector normal to $\Sigma$.

As a last ingredient, these properties can be exploited in combination
with the fact that, by construction, $\mathscr{D}$ is contained in a
larger globally hyperbolic open set ($\mathcal{O}^{\prime}$ in the
notation of the previous section), in order to conclude that
$\varphi_{f}$ can be unambiguously extended to a solution of the very
same equation throughout $\mathcal{O}^{\prime}$.  This can be proved
by recalling that both $\mathscr{D}$ and $\mathcal{O}^{\prime}$ are
globally hyperbolic and by invoking the uniqueness of the causal
propagator.  As a consequence we are entitled to consider the
restriction of $\varphi_{f}$ on $\mathscr{C}^{+}_{p}$ which yields
\begin{equation}
  \label{eq:boundaryf}
  \varphi_{f} \big\vert_{\mathscr{C}^{+}_{p}} \in C^{\infty} (
  \mathscr{C}^{+}_{p} ) \text{.}
\end{equation}

\subsection{Quantum Algebras on $\mathscr{D}$}
\label{ssec:3.1}

After the setup of a classical field theory, we consider a suitable
quantisation scheme to be described as a two-fold process: in a first
step we shall select a suitable algebra of fields which fulfils the
necessary commutation relations and, second, we choose a quantum state
as a functional on this algebra in order to compute the expectation
values of the relevant observables.

Thus let us proceed in logical sequence starting from $\mathscr{D}$,
the bulk spacetime, where we can introduce $\mathscr{F}_{b} (
\mathscr{D} )$ as the subset of sequences with a finite number of
elements lying in
\begin{equation*}
  \bigoplus_{n \geq 0} \otimes_{s}^{n} C^{\infty}_{0} ( \mathscr{D} )
  \text{,} 
\end{equation*}
where $n = 0$ yields $\mathbb{C}$ by definition while
$\otimes_{s}^{n}$ denotes the $n$-fold symmetric tensor product.
According to this definition it is customary to denote a generic $F
\in \mathscr{F}_{b} ( \mathscr{D} )$ as a finite sequence $\{ F_{n}
\}_{n}$ where each $F_{n} \in \otimes_{s}^{n} C^{\infty}_{0} (
\mathscr{D} )$.  We can now promote $\mathscr{F}_{b} ( \mathscr{D} )$
to a topological $^{\ast}$-algebra equipping it with
\begin{itemize}
\item a tensor product $\cdot_{S}$ such that
  \begin{equation*}
    ( F \cdot_{S} G )_{n} = \sum_{p + q = n} \mathcal{S} ( F_{p}
    \otimes G_{q} ) \text{,}
  \end{equation*}
  where $\mathcal{S}$ is the operator which realises total
  symmetrisation;
\item a $^{\ast}$-operation via complex conjugation, \textit{i.e.},
  $\{ F_{n} \}_{n}^{\ast} = \{ \overline{F}_{n} \}_{n}$ for all $F \in
  \mathscr{F}_{b} ( \mathscr{D} )$;
\item the topology induced by the natural one of $\otimes^{n}_{s}
  C^{\infty}_{0} ( \mathscr{D} )$.
\end{itemize}
The above more traditional realisation of $\mathscr{F}_{b} (
\mathscr{D} )$ can be replaced by a novel point of view, thoroughly
developed in \cite{Brunetti:2009aa,Brunetti:0901.2038}.  To be
specific, consider $\mathscr{F}_{b} ( \mathscr{D} )$ as a suitable
subset of the functionals over $C^{\infty} ( \mathscr{D} )$, the
smooth field configurations.  Explicitly, $F \in \mathscr{F}_{b} (
\mathscr{D} )$ yields a functional $F : C^{\infty} ( \mathscr{D} ) \to
\mathbb{R}$ out of the standard pairing between $\otimes^{n}
C^{\infty} ( \mathscr{D} )$ and $\otimes^{n} C^{\infty}_{0} (
\mathscr{D} )$, denoted by $\langle~\thinspace,~\rangle$, via
\begin{equation}
  \label{eq:functional} 
  F ( \varphi ) \doteq \sum_{n = 0}^{\infty} \frac{1}{n!} \langle
  F_{n} , \varphi^{n} \rangle \text{.}
\end{equation}
In order to grasp the connection between the two perspectives, it is
useful to introduce a G\^ateaux derivative,
\begin{equation*}
  F^{(n)} ( \varphi ) h^{\otimes n} =
  \left. \frac{d^{n}}{d\lambda^{n}} F ( \varphi + \lambda h )
  \right\vert_{\lambda = 0} \text{,} \quad \forall h \in C^{\infty} (
  \mathscr{D} ) \text{,}
\end{equation*}
so that $F_{n} \equiv F^{(n)} ( 0 )$.  We shall use alternatively both
pictures in the forthcoming analysis.

The key point in the subsequent quantisation scheme consists in the
modification of the algebraic product $\cdot_{S}$ to yield a new one,
$\star$, which is constructed out of the causal propagator $\Delta$,
unambiguously defined according to \eqref{eq:causalp},
\begin{equation}
  \label{eq:np}
  ( F \star G ) ( \varphi ) = \sum_{n = 0}^{\infty} \frac{i^{n}}{2^{n}
    n!} \big\langle F^{(n)} ( \varphi ) , \Delta^{\otimes n} G^{(n)} (
  \varphi ) \big\rangle \text{,} \qquad \forall F \text{,\thinspace} G
  \in \mathscr{F}_{b} ( \mathscr{D} ) \text{.}
\end{equation}
By direct inspection one realises that $F \star G$ still lies in
$\mathscr{F}_{b} ( \mathscr{D} )$ and, more importantly, that
$(\mathscr{F}_{b} ( \mathscr{D} ) , \star )$ gets the structure of a
$^{\ast}$-algebra under the operation of complex conjugation.

It is important to notice that, up to now, we have not used the
existence of the equation of motion \eqref{eq:eqm} and, therefore, we
can refer to $\mathscr{F}_{b} ( \mathscr{D} )$ as an
\textit{off-shell} algebra.  On the other hand, if one wants to
encompass also the dynamics of the classical system, one needs only to
divide $\mathscr{F}_{b} ( \mathscr{D} )$ by the ideal $\mathscr{I}$
which is the set of elements in $\mathscr{F}_{b} ( \mathscr{D} )$
generated by those of the form $P_{j} F_{n} ( x_{1} , \dots , x_{n}
)$, where $P_{j}$ is the operator in \eqref{eq:eqm} applied to the
$j$-th variable in $F_{n} \in \otimes_{s}^{n} C^{\infty}_{0} (
\mathscr{D} )$.  The outcome is the \textit{on-shell} algebra
$\mathscr{F}_{bo} ( \mathscr{D} ) \doteq \frac{\mathscr{F}_{b} (
  \mathscr{D} )}{\mathscr{I}}$ which inherits the $^{\ast}$-operation
from $\mathscr{F}_{b} ( \mathscr{D} )$ and is nothing but the more
commonly used field algebra.

At this stage, it is important to remark that neither $\mathscr{F}_{b}
( \mathscr{D} )$ nor its on-shell version $\mathscr{F}_{bo} (
\mathscr{D})$ contain all the elements needed to fully analyse the
underlying quantum field theory.  As a matter of fact, objects such as
the components of the stress-energy tensor involve products of fields
evaluated at the same spacetime point, an operation which is \emph{a
  priori} not well-defined due to the distributional nature of the
fields themselves.  To circumvent this obstruction, a standard
procedure calls for the regularisation of these ill-defined objects by
means of a suitable scheme which goes under the name of Hadamard
regularisation.  We shall not dwell here on the technical details but
only highlight some aspects in the appendix.  The interested reader is
referred to \cite{Hollands:2001aa,Hollands:2002ab} for a full account.

In the functional language used before, the problem mentioned above
translates into the impossibility to include in $\mathscr{F}_{b} (
\mathscr{D} )$ objects of the form
\begin{equation*}
  F ( \varphi ) = \int_{\mathscr{D}} d\mu \negthinspace ( g )
  \medspace f ( x ) \varphi^{2} ( x ) \text{,}
\end{equation*}
where $d\mu \negthinspace ( g )$ is the metric-induced volume form,
while $f$ is a test function in $C^{\infty}_{0} ( \mathscr{D} )$ and
$\varphi \in C^{\infty} ( \mathscr{D} )$.  Actually, the star product
\eqref{eq:np} applied to a couple of such fields involves the
ill-defined pointwise product of $\Delta$ with itself.

To solve this problem, we shall follow the line of reasoning of
\cite{Brunetti:2009aa}.  Namely, we introduce a new class of
functionals, $\mathscr{F}_{e} ( \mathscr{D} )$, which have a finite
number of non-vanishing derivatives, the $n$-th of which has to be a
symmetric element of the space of compactly supported distributions
$\mathcal{E}^{\prime} ( \mathscr{D}^{n} )$, whose wave front sets,
moreover, satisfy the following restriction
\begin{equation}
  \label{eq:inters}
  \WF ( F_{n} ) \cap \big\{ \big( \mathscr{D} \times
  \overline{V}^{\thinspace +} \big)^{n} \cup \big( \mathscr{D} \times
  \overline{V}^{\thinspace -} \big)^{n} \big\}  = \emptyset \text{,}
\end{equation}
where $\overline{V}^{\thinspace \pm}$ corresponds to the forward and
to the backward causal cone in the tangent space, respectively.  The
closure symbol indicates that we also include the tip of the cone in
the set of future or past directed causal vectors.

We can make $\mathscr{F}_{e} ( \mathscr{D} )$ a $^{\ast}$-algebra if
we extend naturally the $^{\ast}$-operation of $\mathscr{F}_{b} (
\mathscr{D} )$ and endow it with a new product, $\star_{H}$, whose
well-posedness was first proved in
\cite{Brunetti:1996aa,Brunetti:2000aa,Hollands:2001aa,Hollands:2002ab}.
The explicit form is realised as
\begin{equation}
  \label{eq:starproduct} 
  ( F \star_{H} G ) ( \varphi ) = \sum_{n=0}^{\infty} \frac{1}{n!}
  \big\langle F^{(n)} ( \varphi ) , H^{\otimes n} G^{(n)} ( \varphi )
  \big\rangle \text{,}
\end{equation}
where $H \in \mathcal{D}^{\prime} ( \mathscr{D}^{2} )$ is the
so-called Hadamard bi-distribution.  We shall briefly introduce and
discuss it in the appendix, but, for our purposes, it is important to
recall that, on the one hand, it satisfies the microlocal spectrum
condition, hence yielding a natural substitute for the notion of
positivity of energy out of its wave front set, while on the other
hand it suffers from an ambiguity.  At the level of the integral
kernel, only the antisymmetric part of the Hadamard bi-distribution is
fixed to $i/2$ times the causal propagator $\Delta$.  Also the
singular structure is unambiguously determined by the choice of the
background.  Otherwise there always exists the freedom to add a smooth
symmetric function which, in our scenario, means that, if $H$,
$H^{\prime}$ are two Hadamard distributions, then the integral kernel
of $H - H^{\prime}$ is a symmetric element of $C^{\infty} (
\mathscr{D}^{2} )$.  Yet, as far as the algebra is concerned, this
freedom boils down to an algebraic isomorphism
$\mathfrak{i}_{H^{\prime} , H} : ( \mathscr{F}_{e} ( \mathscr{D} ) ,
\star_{H}) \to ( \mathscr{F}_{e} ( \mathscr{D} ) ,
\star_{H^{\prime}})$, namely
(\textit{cf.}~\cite{Hollands:2001aa,Brunetti:0901.2038})
\begin{equation}
  \label{eq:isomorphism}
  \begin{split}
    \mathfrak{i}_{H^{\prime} , H} &= \alpha_{H^{\prime}} \circ
    \alpha_{H}^{-1} \text{,} \\
    \alpha_{H} ( F ) & \doteq \sum_{n = 0}^{\infty} \frac{1}{n!}
    \big\langle H^{\otimes n} , F^{(2n)} \big\rangle \text{.}
  \end{split}
\end{equation}

As for the algebra generated by compactly supported smooth functions,
also the extended one, $\mathscr{F}_{e}$, has its on-shell
counterpart, $\mathscr{F}_{eo}$, constructed from the quotient with
the ideal generated by the equation of motion applied to the elements
of $\mathscr{F}_{e}$.

One of the advantages of the formalism employed is the possibility to
easily transcribe the overall construction in terms of categories,
hence yielding a crystal-clear mathematical picture of the relevant
structures and their relations.  This was first advocated and utilised
in the seminal paper \cite{Brunetti:2003aa}, where the principle of
general local covariance was formulated in this language to which we
shall also stick.  In particular, we shall now recast the above
discussion in this different perspective, while the actual relation
with \cite{Brunetti:2003aa} will only later be outlined in
Section~\ref{sec:4}.  Hence, we shall use the following categories:
\begin{itemize}
\item[$\mathsf{DoCo}$:] The objects are defined as follows: For every
  spacetime $M$, as in the previous section, we consider the oriented
  and time-oriented double cones $\mathscr{D} ( p , q)$ with the
  property that there exists a normal neighbourhood $\mathcal{O}_{p}
  \subset M$ centred in $p$ that contains $\mathscr{D} ( p , q )$.  Hence, 
  an object is a triple $\mathscr{D} \equiv [ \mathscr{D} (
  p , q ) , \mathcal{O}_{p}, e]$.  Recall that since both $\mathcal{O}_{p}$ and the double cones are
  globally hyperbolic, the choice of time and space orientation is
  always possible.  The morphisms are the maps
  $\imath_{e , e^{\prime}} : \mathcal{O}_{p} \to
  \mathcal{O}_{p^{\prime}}$ introduced in
  \eqref{eq:morphisms} such that $\imath_{e , e^{\prime}}
  \big\vert_{\mathscr{D}} \big( \mathscr{D} ( p , q ) \big) \subset
  \mathscr{D}^{\prime} ( p^{\prime} , q^{\prime})$. Although the subscript which refers
to $e$ and $e^\prime$ is a small abuse of notation, we feel that its presence might help the reader
to focus on the ingredients here at play. The composition
  rule for the morphisms is defined in terms of the composition of the
  maps $\imath_{e , e^{\prime}}$ and hence the associativity derives
  from the associativity of the composition.  Notice that, as per
  \eqref{eq:morphisms}, an arrow between two objects exists only if
  the source double cone is sufficiently small so that its image under
  $\exp^{-1}_{p}$ is contained in the domain of the definition of
  $\exp_{p^{\prime}}$.  This caveat does not spoil the associativity
  property of the composition of arrows.
\item [$\mathsf{DoCo}_{iso}$:] This is the subcategory of
  $\mathsf{DoCo}$ obtained by keeping the same objects but restricting
  the possible morphisms of $\mathsf{DoCo}$ to those which are
  isometric embeddings.
\item[$\mathsf{Alg}_{i}$:] The objects are unital, topological,
  $^{\ast}$-algebras $\mathscr{F}_{i} ( \mathscr{D} )$ with $i = b$,
  $bo$, constructed as above, while for $i = e$, $eo$, we further
  restrict the objects to the equivalence classes generated by
  identifying extended algebras which are isomorphic under the map
  \eqref{eq:isomorphism}.  The morphisms are equivalence classes of injective,
  unit-preserving, $^{\ast}$-homomorphisms.
\end{itemize}

Since the key ingredients to construct both $\mathscr{F}_{b} (
\mathscr{D} )$ and $\mathscr{F}_{bo} ( \mathscr{D} )$ are just the
causal propagator $\Delta$ from \eqref{eq:causalp} and the operator
\eqref{eq:eqm} realising the equations of motion, their uniqueness in
any $\mathscr{D}$ suggests that the association of a suitable algebra,
\begin{equation}
  \label{eq:funct1}
  \mathscr{F}_{i} : \mathscr{D} \to \mathscr{F}_{i} ( \mathscr{D} )
  \text{,} \qquad i = b \text{,~} bo \text{,}
\end{equation}
with each double cone $\mathscr{D}$ can be promoted to a functor
between $\mathsf{DoCo}_{iso}$ and $\mathsf{Alg}_{i}$.  To this end, in
order to define the action of $\mathscr{F}_{i}$ on morphisms of
$\mathsf{DoCo}_{iso}$, notice that $\mathscr{F}_{i} ( \mathscr{D} )$
is an algebra generated by smooth and compactly supported functions on
$\mathscr{D}$, hence $\mathscr{F}_{i} ( \imath_{e , e^{\prime}})$ is
just the injective $^{\ast}$-homomorphism which associates with every
element in $\mathscr{F}_{i} ( \mathscr{D} )$ its image under
$\imath_{e , e^{\prime}}$.  Furthermore, $\mathscr{F}_{i}$ defined in
that way enjoys the covariance property and maps $\id_{\mathscr{D}}$,
the identity of $\mathscr{D}$ in $\mathsf{DoCo}_{iso}$, to
$\id_{\mathscr{F}_{i} ( \mathscr{D} )}$, the identity of
$\mathscr{F}_{i} ( \mathscr{D} )$ in $\mathsf{Alg}_{i}$,
\textit{i.e.},
\begin{equation*}
  \mathscr{F}_{i} ( \imath_{e , e^{\prime}} ) \circ \mathscr{F}_{i} (
  \tilde{\imath}_{e^\prime , \tilde{e}^{\prime}} ) = \mathscr{F}_{i}
  ( \imath_{e , e^{\prime}} \circ \tilde{\imath}_{e^\prime ,
    \tilde{e}^{\prime}} ) \qquad \text{and} \qquad \mathscr{F}_{i} (
  \id_{\mathscr{D}} ) = \id_{\mathscr{F}_{i} ( \mathscr{D} )} \text{.}
\end{equation*}
Notice that, in the case of the on-shell algebra, the equation of
motion is left unchanged by any morphism in $\mathsf{DoCo}_{iso}$.

It is also important to note that singling out extended algebras which
are related via \eqref{eq:isomorphism} ensures that the ambiguity in
the choice of the Hadamard bi-distribution does not spoil the
well-posedness of \eqref{eq:funct1} when $i = e$.  All these
assertions can be proved noticing that, due to the discussion
presented after \eqref{eq:morphisms}, the category
$\mathsf{DoCo}_{iso}$ is just a subcategory of the category of local
manifolds introduced in \cite{Brunetti:2003aa} where similar results
are discussed.

It would be desirable to extend the functor \eqref{eq:funct1} to the
category $\mathsf{DoCo}$ that has a larger group of morphisms.
Unfortunately, this is not straightforward and, actually, not even
possible.  If we consider two generic globally hyperbolic regions
$\mathscr{D}$ and $\mathscr{D}^{\prime}$ in $\mathsf{DoCo}$, related
by $\imath_{e , e^{\prime}}$ as in \eqref{eq:morphisms}, we can draw
the following diagram,
\begin{equation}
  \label{eq:opendiagram}
  \begin{CD}
    \mathscr{D} @>{\mathscr{F}}>>  \mathscr{F}_{i} ( \mathscr{D} ) \\
    @V{\imath_{e , e^{\prime}}}VV \\
    \mathscr{D}^{\prime} @>>{\mathscr{F}}> \mathscr{F}_{i} (
    \mathscr{D}^{\prime} )
  \end{CD}
  \qquad \raisebox{-2\baselineskip}{\text{.}}
\end{equation}
To have a functor between $\mathsf{DoCo}$ and some $\mathsf{Alg}_{i}$
requires existence of a well-defined morphism $\mathscr{F} ( \imath_{e
  , e^{\prime}})$ between $\mathscr{F}_{i} ( \mathscr{D} )$ and
$\mathscr{F}_{i} ( \mathscr{D}^{\prime} )$.  But this is not possible
since, in general, $\imath_{e , e^{\prime}}$ is not an isometry and,
thus, it does neither map solutions of \eqref{eq:eqm} in $\mathscr{D}$
into those of the same equation (but out of a different metric) in
$\mathscr{D}^{\prime}$, nor does it preserve the causal propagator.
Hence it spoils the $\star$-operation and does not preserve the
singular structure of the Hadamard bi-distribution which only depends
upon the underlying geometry.

As an aside, a positive answer to the present question will only be
possible considering the off-shell classical $^{\ast}$-algebras $(
\mathscr{F}_{b} , \cdot_{S} )$, but, as soon as quantum algebras are
employed, the situation looks grim.  Yet it is possible to circumvent
this obstruction making profitable use of the geometric properties of
the portion of the future directed light cone in any $\mathscr{D}$ in
order both to set up a bulk-to-boundary correspondence and to later
compare the outcome on the boundary of different spacetimes.

\subsection{Quantum Field Theory on the Boundary}
\label{ssec:3.2}

In order to fulfil our goals, it is mandatory as a first step to
understand how to construct a full-fledged quantum field theory on the
light cone and this will be the main aim of this subsection.  Our
procedure derives from the experience gathered in similar scenarios
where a field theory on a null surface was constructed such as,
\emph{e.g.}, in
\cite{Dappiaggi:2006aa,Dappiaggi:2009aa,Dappiaggi:2009ab,Moretti:2004aa}
(see also \cite{Herdegen:2008aa,Schroer:0905.4435,Chmielowiec:2009aa}
for further analyses in similar contexts).

Therefore, following the same philosophy as in these papers, we shall
first show that it is possible to assign to the boundary a natural
field algebra and that there is a ``natural'' choice for the relevant
quantum state.  To this avail, in this subsection, we shall consider
the cone as an abstract manifold on its own, not regarding it as a
particular portion of the boundary of a specific globally hyperbolic
double cone $\mathscr{D}$, since the connection with the bulk theory
will be presented only later.  Nevertheless, we have to keep in mind
that the algebra to be constructed has to be large enough in order to
contain the images of some suitable projections of all the elements of
the algebra in the bulk $\mathscr{D}$.  This will be the most delicate
point of the whole construction because it is not sufficient to
consider an algebra generated by compactly supported data on the cone;
we have to extend this set to more generic elements.

Let us start with the introduction of three distinct sets in
$\mathbb{R} \times \mathbb{S}^{2} \subset \mathbb{R}^{4}$ relevant for the
following construction, namely, employing the standard coordinates, we
define
\begin{equation}
  \label{eq:cone1}
  \mathscr{C}^{+}_{p} = \big\{ ( V , \theta , \varphi ) \in \mathbb{R}
  \times \mathbb{S}^{2} \medspace \vert \medspace V \in I \subset
  \mathbb{R} \text{,} \thickspace ( \theta , \varphi ) \in
  \mathbb{S}^{2} \big\} \text{,}
\end{equation}
where $I$ is the open interval $\big( 0 , V_{0} ( \theta , \varphi )
\big)$ with a positive, bounded smooth function $V_{0} ( \theta ,
\varphi )$ on the sphere.  The other two regions will be denoted
$\mathscr{C}_{p}$ and $\mathscr{C}$, respectively, where the
coordinate $V$ is allowed to extend over $( 0 , \infty )$ or the full
real line, so that $\mathscr{C}^{+}_{p} \subset \mathscr{C}_{p}
\subset \mathscr{C}$.  We stress that, with a slight abuse of
notation, we employ the symbol $\mathscr{C}^{+}_{p}$ as in the
previous sections although we are not referring to an actual cone
since, ultimately, \eqref{eq:cone1} will indeed coincide with
$J^{+}_{p} \cap \partial {\mathscr{D}}$, if we employ the same
conventions and nomenclatures as in the preceding analysis.
Furthermore notice that $\mathscr{C}$ is not completely independent of
$p$ which still is required to be a point in this set.  Yet, to avoid
worsening an already cumbersome notation, we leave such a dependence
implicit.

As a natural next step, we need to identify a suitable space of
functions on the boundary and, to this avail, viewing $\mathscr{C}_{p}$
immersed in $\mathbb{R}^{4}$, we define
\begin{equation}
  \label{eq:mS}
  \mathscr{S} ( \mathscr{C}_{p} ) \doteq \Big\{ \psi \in C^{\infty} (
  \mathscr{C}_{p} ) \medspace \vert \medspace \psi = h f
  \big\vert_{\mathscr{C}_{p}} \text{,} \thickspace f \in
  C^{\infty}_{0} ( \mathbb{R}^{4} ) \text{~and~} h \in C^{\infty} (
  \mathscr{C}_{p} ) \Big\} \text{,}
\end{equation}
where $h$ vanishes uniformly on $\mathbb{S}^{2}$, as $V \to 0$, while
each derivative along $V$ tends to a constant uniformly on
$\mathbb{S}^{2}$.  As far as this subsection is concerned we can
safely choose $h$ to be equal to $V$.  Furthermore, $\mathscr{S} (
\mathscr{C}_{p} )$ turns out to be a symplectic space when endowed
with the following strongly non-degenerate symplectic form,
\begin{equation}
  \label{eq:symplecticform}
  \sigma_{\mathscr{C}} ( \psi , \psi^{\prime} ) \doteq
  \int_{\mathscr{C}_{p}} \Big[ \psi \frac{d\psi^{\prime}}{dV} -
  \frac{d\psi}{dV} \psi^{\prime} \Big] \medspace dV \wedge
  d\mathbb{S}^{2} \text{,} \quad \forall \psi \text{,} \psi^{\prime}
  \in \mathscr{S} ( \mathscr{C}_{p} ) \text{,}
\end{equation}
where $d\mathbb{S}^{2}$ is the standard measure on the unit
$2$-sphere.

The reason for such an apparently strange choice of $\mathscr{S} (
\mathscr{C}_{p} )$ is the later need to relate the theory on these
sets to the one in the bulk of a double cone.  Hence the most natural
choice of compactly supported smooth functions on $\mathscr{C}_{p}$
would not fit into the overall picture since a general solution of the
Klein-Gordon equation \eqref{eq:eqm} with smooth compactly supported
initial data on some $\mathscr{D}$ would propagate on the light cone
$\mathscr{C}_{p}$ to a function which is also supported on the tip.
This point corresponds to $V = 0$ in the above picture and, thus, it
does not lie in $\mathscr{C}_{p}$.  On the contrary, we shall show in
the next subsection that \eqref{eq:mS} is indeed the natural
counterpart on the boundary which arises from the set of solutions of
\eqref{eq:eqm}.

In order to introduce the relevant algebra of observables, we follow
the same philosophy as in Subsection~\ref{ssec:3.1}, introducing
$\mathscr{A}_{b} ( \mathscr{C}_{p} )$, whose generic element $F^{\prime}$ is a
sequence $\{ F^{\prime}_{n} \}_{n}$ with a finite number of elements in
\begin{equation}
  \label{eq:boundalg}
  \bigoplus_{n \geq 0} \otimes_{s}^{n} \mathscr{S} ( \mathscr{C}_{p} )
  \text{,}
\end{equation}
where $\otimes_{s}^{n}$ again denotes the symmetrised $n$-fold tensor
product and the first term in the sum is $\mathbb{C}$.  Notice that
the $^{\prime}$-superscript is introduced in this subsection in order
to avoid a potential confusion with the similar symbols used for the
counterpart in the bulk.  In order to promote \eqref{eq:boundalg} to a
full topological $^{\ast}$-algebra, we have to endow it with
\begin{itemize}
\item a $^{\ast}$-operation which is the complex conjugation,
  \textit{i.e.}, $\{ F^{\prime}_{n} \}_{n}^{\ast} = \{
  \overline{F_{n}^{\prime}} \}_{n}$ for all $F^{\prime} \in
  \mathscr{A}_{b} ( \mathscr{C}_{p} )$;
\item multiplication of elements such that for any $F^{\prime}$,
  $G^{\prime} \in \mathscr{A}_{b} ( \mathscr{C}_{p} )$,
  \begin{equation*}
    ( F^{\prime} \cdot_{S} G^{\prime} )_{n} = \sum_{p + q = n}
    \mathcal{S} ( F^{\prime}_{p} \otimes G^{\prime}_{q} ) \text{;}
  \end{equation*}
\item the topology induced by the natural topology of $\mathscr{S} (
  \mathscr{C}_{p} )$, namely the topology of smooth functions on
  $\mathscr{C}_{p}$.
\end{itemize}
Although well-defined, this algebra is not suited to be put in
relation to data in the bulk and, thus, the above product has to be
deformed once more.  To this avail, we employ the functional point of
view as in \eqref{eq:functional}, \textit{i.e.}, if $X \doteq \big\{
\Phi \in C^{\infty} ( \mathscr{C}_{p} ) \medspace \vert \medspace
V^{-1} \Phi \in C^{\infty} ( \mathscr{C} ) \big\}$, then $F^{\prime}
\in \mathscr{A}_{b} ( \mathscr{C}_{p} )$ yields a functional
$F^{\prime} : X \to \mathbb{R}$ out of $\langle~\thinspace,~\rangle$,
the pairing between $\otimes^{n} X$ and $\otimes^{n} \mathscr{S} (
\mathscr{C}_{p} )$,
\begin{equation*}
  F^{\prime} ( \Phi ) = \sum_{n = 0}^{\infty} \frac{1}{n!} \langle
  F^{\prime}_{n} , \Phi^{n} \rangle \text{,} \quad \forall \Phi \in X
  \text{.}
\end{equation*}
By direct inspection of the definition of $\mathscr{S} (
\mathscr{C}_{p} )$ in \eqref{eq:mS}, one notices that
$\langle~\thinspace,~\rangle$ is nothing but the standard inner
product on $( 0 , \infty ) \times \mathbb{S}^{2}$ between compactly
supported functions and smooth ones, taken with respect to the measure
$dV \wedge d\mathbb{S}^{2}$.

Although the theory on $\mathscr{C}_{p}$ has no equation of motion built
in and, hence, no causal propagator such as \eqref{eq:causalp}, we can
nonetheless introduce a new $\star_{B}$-product on
$\mathscr{A}_{b} ( \mathscr{C}_{p} )$, namely
\begin{equation}
  \label{eq:starproductbordo}
  ( F^{\prime} \star_{B} G^{\prime} ) ( \Phi ) \doteq \sum_{n =
    0}^{\infty} \frac{i^n }{2^n n!} \big\langle F^{\prime}_{n} ( \Phi
  ) , \Delta_{\sigma_\mathscr{C}}^{n} G^{\prime}_{n} ( \Phi )
  \big\rangle \text{,} \quad \forall \Phi \in X \text{,}
\end{equation}
where $\Delta_{\sigma_\mathscr{C}}$ is the integral kernel of
\eqref{eq:symplecticform}, \textit{i.e.},
\begin{equation}
  \label{eq:symplectickernel}
  \Delta_{\sigma_\mathscr{C}} \big( ( V , \theta , \varphi ) , (
  V^{\prime} , \theta^{\prime} , \varphi^{\prime} ) \big) = -
  \frac{\partial^{2}}{\partial V \partial V^{\prime}} \sign ( V -
  V^{\prime}) \medspace \delta ( \theta , \theta^{\prime} ) \text{,}
\end{equation}
where $\delta ( \theta , \theta^{\prime} )$ stands for $\delta (
\theta - \theta^{\prime} ) \delta ( \varphi - \varphi^{\prime} )$ and
the derivatives have to be taken in the weak sense.  Notice that
$\Delta_{\sigma_\mathscr{C}}$ is defined as a distribution on
$C^{\infty}_{0} ( \mathscr{C}_{p}^{2})$ which, by direct inspection,
can be extended also to $\mathscr{S}^{2} ( \mathscr{C}_{p}^{2})$.
Finally, $\star_{B}$ is well-defined because only a finite number of
elements appear in the sum on the right-hand side of
\eqref{eq:symplectickernel} and, thus, convergence is not an issue
here.  We can hence conclude this subsection with a proposition whose
proof follows from the preceding discussion.
\begin{proposition}
  The pair $\big( \mathscr{A}_{b} ( \mathscr{C}_{p} ) , \star_{B}
  \big)$ equipped with the $^{\ast}$-operation introduced above is a
  well defined $^{\ast}$-algebra.
\end{proposition}
In the next section we shall discuss the form of a certain class of
quantum states on this algebra to eventually use them to extend the
boundary algebra of observables analogous to the procedure on the
bulk.

\subsection{Natural Boundary States}
\label{ssec:3.3}

The next step in our construction is the introduction of an extended
algebra on the boundary, but, in the present scenario, there is no
standard definition of an Hadamard state or of a bi-distribution; and
this lack of a class of \textit{a priori} physically relevant states
hinders an imitative repetition of the use of the function $H$ as in
\eqref{eq:starproduct}.  Therefore, a natural bi-distribution on
$\mathscr{C}^{+}_{p}$ is needed and, for this purpose, our choice is
the following weak limit,
\begin{equation}
  \label{eq:state}
  \omega \big( ( V , \theta , \varphi ) , ( V^{\prime} ,
  \theta^{\prime} , \varphi^{\prime} ) \big) \doteq - \frac{1}{\pi}
  \lim_{\epsilon \to 0^{+}} \frac{1}{( V - V^{\prime} - i \epsilon
    )^{2}} \medspace \delta ( \theta , \theta^{\prime} ) \text{,}
\end{equation}
which has the advantage of being at the same time a well-defined
element of $\mathcal{D}^{\prime} ( \mathscr{C}^{2} )$, where
$\mathscr{C} \sim \mathbb{R} \times \mathbb{S}^{2}$ and
$\mathcal{D}^{\prime}$ denotes the space of distributions over the
test functions in $C^{\infty}_{0} ( \mathscr{C} )$.

Such an expression already appeared in different, albeit related
scenarios where a scalar quantum field theory was studied
\cite{Dappiaggi:2006aa,Dappiaggi:2009ab,Dappiaggi:0907.1034,%
  Moretti:2004aa,Kay:1991aa,Schroer:0905.4435,Sewell:1982aa}.  It is
important to remark that in the first two of these papers,
\eqref{eq:state} was actually used as the building block to construct
a quasi-free pure state for a scalar field theory built on a
three-dimensional null cone which represented the conformal boundary
of a suitable class of spacetimes.  In all these cases, the particular
geometric structure as well as the presence of a particular symmetry
group entails that \eqref{eq:state} fulfils suitable uniqueness
properties and, furthermore, gives rise to a full-fledged Hadamard
state in the bulk.  Therefore we shall employ the above expression as
the natural candidate bi-distribution on the boundary, proving in the
next sections that, when we realise $\mathscr{C}^{+}_{p}$ as part of
the boundary of a double cone, we can also construct a physically
meaningful state on the bulk $\mathscr{D}$ out of \eqref{eq:state}.

The above bi-distribution can be read as a functional $\omega :
\mathscr{S} ( \mathscr{C}_{p} ) \times \mathscr{S} ( \mathscr{C}_{p})
\to \mathbb{R}$, but it is actually much more convenient to recall
that $\mathscr{C}_{p} \subset \mathscr{C}$.  Within this perspective,
since $\mathscr{C}$ is topologically $\mathbb{R} \times
\mathbb{S}^{2}$ and each element in $\mathscr{S} ( \mathscr{C}_{p} )$,
together with the $V$-derivative, also lies in $L^{2} ( \mathscr{C} ,
dV \wedge d\mathbb{S}^{2})$, the following expression for the
$2$-point function is meaningful,
\begin{equation}
  \label{eq:2-pt}
  \omega ( \psi , \psi^{\prime} ) = - \frac{1}{\pi} \lim_{\epsilon \to
    0^{+}} \int_{\mathbb{R}^{2} \times \mathbb{S}^{2}} dV dV^{\prime}
  d\mathbb{S}^{2} ( \theta , \varphi ) \medspace \frac{\psi ( V ,
    \theta , \varphi ) \psi^{\prime} ( V^{\prime} , \theta , \varphi
    )}{( V - V^{\prime} - i \epsilon )^{2}} \text{,}
\end{equation}
where the delta function over the angular coordinates is already
integrated out.  The distribution \eqref{eq:2-pt} satisfies a suitable
continuity condition, as for example shown in \cite{Moretti:2008aa},
\begin{equation}
  \label{eq:2pcontinuity}
  \lvert \omega ( \psi , \psi^{\prime} ) \rvert \leqslant C \big(
  \lVert \psi \rVert_{L^{2}} \lVert \partial_{V} \psi \rVert_{L^{2}}
  \big) \big( \lVert \psi^{\prime} \rVert_{L^{2}} +
  \lVert \partial_{V} \psi^{\prime} \rVert_{L^{2}} \big) < \infty
  \text{,}
\end{equation}
which allows for the extension of $\omega$ to the space of
square-integrable functions whose derivative along the $V$-coordinate
is also an element of $L^{2}$.  Furthermore, this also entails the
possibility to perform a Fourier-Plancherel transform along $V$
(\textit{cf.} Appendix C in \cite{Moretti:2008aa}) resulting in a much
more manageable form for \eqref{eq:2-pt},
\begin{equation}
  \label{eq:L2integral}
  \omega ( \psi , \psi^{\prime} ) = \int_{\mathbb{R} \times
    \mathbb{S}^{2}}  dk \thinspace d\mathbb{S}^{2} ( \theta , \varphi
  ) \thickspace 2 k \medspace \Theta ( k ) \medspace
  \overline{\widehat{\psi} ( k , \theta , \varphi )} \thinspace
  \widehat{\psi}^{\prime} ( k , \theta , \varphi ) \medspace \text{,}
\end{equation}
where $\Theta ( k )$ is the step function equal to $1$ if $k \geqslant
0$ and $0$ otherwise.  It should be stressed that the presence of
$\Theta ( k )$ corresponds to the physical intuition of taking only
positive frequencies, also because, on the cone, the only causal
directions are the lines at constant angular variables.  Under special
circumstances this idea has a clear connection with the geometrical
bulk data as well as with the Hadamard property of a bulk state
constructed out of \eqref{eq:state} (\textit{cf.} for example
\cite{Moretti:2006aa,Moretti:2008aa,Dappiaggi:2009ab}).

As a last step in this subsection, we underline that the above
analysis entails two relevant remarks.  The first one concerns the
wave front set of the bi-distribution $\omega$ on $\mathscr{C}^{2}
\sim ( \mathbb{R} \times \mathbb{S}^{2} )^{2}$.  This was already
studied in Lemma~4.4. of \cite{Moretti:2008aa} yielding
\begin{equation}
  \label{eq:WFomega}
  \WF ( \omega ) \subseteq A \cup B \text{,}
\end{equation}
where
\begin{multline}
  \label{eq:WFA}
  A = \big\{ \big( ( V , \theta , \varphi , \zeta_{V} ,
  \zeta_{\theta}, \zeta_{\varphi} ) , ( V^{\prime} , \theta^{\prime} ,
  \varphi^{\prime} , \zeta_{V^{\prime}} , \zeta_{\theta^{\prime}} ,
  \zeta_{\varphi^{\prime}} ) \big) \in ( T^{\ast} \mathscr{C} )^{2}
  \setminus \{0\} \thickspace \vert \\
  V = V^{\prime} \text{,~} \theta = \theta^{\prime} \text{,~} \varphi
  = \varphi^{\prime} \text{,~} 0 < \zeta_{V} = -\zeta_{V^{\prime}}
  \text{,~} \zeta_{\theta} = -\zeta_{\theta^{\prime}} \text{,~}
  \zeta_{\varphi} = -\zeta_{\varphi^{\prime}} \big\}
\end{multline}
and
\begin{multline}
  \label{eq:WFB}
  B = \big\{ \big( ( V , \theta ,\varphi , \zeta_{V} , \zeta_{\theta}
  , \zeta_{\varphi} ) , ( V^{\prime} , \theta^{\prime} ,
  \varphi^{\prime} , \zeta_{V^{\prime}} , \zeta_{\theta^{\prime}} ,
  \zeta_{\varphi^{\prime}} ) \big) \in ( T^{\ast} \mathscr{C} )^{2}
  \setminus \{0\} \thickspace \vert \\
  \theta = \theta^{\prime} \text{,~} \varphi = \varphi^{\prime}
  \text{,~} \zeta_{V} = \zeta_{V^{\prime}} = 0 \text{,~}
  \zeta_{\theta} = -\zeta_{\theta^{\prime}} \text{,~} \zeta_{\varphi}
  = -\zeta_{\varphi^{\prime}} \big\} \text{.}
\end{multline}
Although, at this stage, this result is only an aside, it will play a
pivotal role in the discussion of Subsections~\ref{ssec:3.4} and
\ref{ssec:3.5}.  In particular, if we recall that $\mathscr{C}_{p}
\subset \mathscr{C}$, it turns out that the wave front set of $\omega$
on $C^{\infty}_{0} ( \mathscr{C}_{p}^{2} )$ can only be smaller or
equal to $A \cup B$, and actually it corresponds to $A \cup B$
restricted to $( T^{\ast} \mathscr{C}_{p} )^{2}$.

As a second remark, note that it is possible to construct a new
algebra, say $\mathscr{A}_{b} ( \mathscr{C} )$ on the full
$\mathscr{C}$ starting from \eqref{eq:mS} and considering the set
$L^{2} ( \mathscr{C} , dV \wedge d\mathbb{S}^{2} )$ in place of
$\mathscr{S} ( \mathscr{C}_{p} )$, while keeping the same
$^{\ast}$-operation and composition rule.  On the one hand, it is
straightforward, that $\mathscr{A}_{b} ( \mathscr{C}_{p} )$ is a
$^{\ast}$-subalgebra of $\mathscr{A}_{b} ( \mathscr{C} )$ while, on
the other hand, we can see that the two-point function $\omega$ as in
\eqref{eq:state} can be used as a building block of a quasi-free state
for $\mathscr{A}_{b} ( \mathscr{C} )$.  Hence the same conclusion can
be drawn for $\mathscr{A}_{b} ( \mathscr{C}_{p} )$ since, by
construction, the antisymmetric part of $\omega$ is equal to
$\frac{i}{2} \Delta_{\sigma}$.  The only possible issue is positivity,
but this is solved by direct inspection of \eqref{eq:L2integral} whose
right-hand side is manifestly greater than $0$ once $\psi =
\psi^{\prime}$.  It is important to point out, for the sake of
completeness, that an almost identical analysis appears in
\cite{Dappiaggi:2006aa,Dappiaggi:2009ab}, though performed at the
level of Weyl algebras. In summary, we get
\begin{proposition}
  \label{Pro:stateomega}
  The Gaussian (quasi-free) state constructed out of the distribution
  $\omega$ enjoys the following properties:
  \begin{enumerate}
  \item It is a well-defined algebraic state on $\mathscr{A}_{b} (
    \mathscr{C}_{p})$ and on $\mathscr{A}_{b} ( \mathscr{C} )$.
  \item It is a vacuum with respect to $\partial_{V}$, in the sense
    given in \cite{Sahlmann:2000aa}.
  \item It is invariant under the change of the local frame, hence
    invariant under the action of $SO_{0} (1,3)$.
  \end{enumerate}
\end{proposition}
\begin{proof}
  The first point can be analysed by checking linearity, positivity
  and normalisability of the state on $\mathscr{A}_{b} (
  \mathscr{C}_{p} )$.  Since the state is quasi-free, it is enough to
  examine these properties for the functional $\omega$ on $\mathscr{S}
  ( \mathscr{C}_{p} ) \times \mathscr{S} ( \mathscr{C}_{p} )$ where
  they follow from the previous discussion.

  The proof of the second point derives from the observation that
  $\omega$, as a state on $\mathscr{A}_{b} ( \mathscr{C} )$, is
  invariant under translations and from the fact that the
  Fourier-Plancherel transform of the integral kernel of $\omega$
  along the $V$-direction only contains positive frequencies, as is
  clear from \eqref{eq:L2integral}.

  The third point can be proved recalling a result in
  \cite{Dappiaggi:2006aa}, namely, $\omega$ on $\mathscr{A}_{b} (
  \mathscr{C} )$ is invariant under the action of an
  infinite-dimensional group, the so-called Bondi-Metzner-Sachs group
  (BMS).  In short, if one switches from the coordinates $( V , \theta
  , \varphi )$ to $( V , z , \bar{z} )$ obtained out of a
  stereographic projection, the BMS maps
  \begin{equation*}
    \begin{cases}
      z \mapsto z^{\prime} = \Lambda ( z ) \doteq \frac{a z + b}{c z +
        d} \text{,} \quad a d - b c = 1 \text{,} \quad a \text{,~} b
      \text{,~} c \text{,~} d \in \mathbb{C} \text{,} \\
      V \mapsto V^{\prime} = K_{\Lambda} ( z , \bar{z} ) ( V + \alpha
      ( z , \bar{z} ) ) \text{,}
    \end{cases}
  \end{equation*}
  where $\bar{z}$ transforms as the complex conjugate of $z$,
  $\alpha ( z , \bar{z} ) \in C^{\infty} ( \mathbb{S}^{2} )$
  and
  \begin{equation*}
    K_{\Lambda} ( z , \bar{z} ) \doteq \frac{1 + \abs{z}^{2}}{\abs{a z
        + d}^{2} - \abs{b z + c}^{2}} \text{.}
  \end{equation*}
  Hence, by direct inspection of the above formulae, the BMS group is
  seen to be the regular semidirect product $SL (2,\mathbb{C}) \rtimes
  C^{\infty} ( \mathbb{S}^{2} )$.  Most notably one observes that
  there exists a proper subgroup which is homomorphism to $SO (3,1)$
  and thus the state $\omega$ turns out to be invariant under the
  group sought-for.  \qed
\end{proof}

\subsection{Extended Algebra on the Boundary}
\label{ssec:3.4}

In the previous subsection, we introduced the boundary algebra
together with a suitable notion of $\star$-product, but this is still
not sufficient to intertwine the boundary data with those on the bulk
because we lack a counterpart for the extended algebra of observables
on $\mathscr{C}_{p}$.  Yet, due to the results of the last subsection,
we have all the ingredients to construct it.

As a starting point, we define the building block of the extended
algebra as follows.
\begin{definition}
  \label{Def:testdistributions}
  We call $\mathcal{A}^{n}$ the set of elements $F^{\prime}_{n} \in
  \mathcal{D}^{\prime} ( \mathscr{C}^{n}_{p} )$ that fulfil the
  following properties:
  \begin{enumerate}
  \item \textbf{Compactness}: The $F^{\prime}_{n}$ are compact towards
    the future, \textit{i.e.}, the support of $F^{\prime}_{n}$ is
    contained in a compact subset of $\mathscr{C}^{n} \sim (
    \mathbb{R} \times \mathbb{S}^{2} )^{n}$.
  \item \textbf{Causal non-monotonic singular directions}: The wave
    front set of $F^{\prime}_{n}$ contains only causal non-monotonic
    directions which means that
    \begin{equation}
      \label{eq:WFS} 
      \WF ( F^{\prime}_{n} ) \subseteq W_{n} \doteq \big\{ ( x , \zeta
      ) \in ( T^{\ast} \mathscr{C}_{p} )^{n} \setminus \{0\} \medspace
      \vert \medspace ( x , \zeta ) \not\in \overline{V}_{n}^{+} \cup
      \overline{V}_{n}^{-} \text{,~} ( x , \zeta ) \not\in S_{n}
      \big\} \text{,}
    \end{equation}
    where $( x , \zeta ) \equiv ( x_{1} , \dotsc , x_{n} , \zeta_{1} ,
    \dotsc , \zeta_{n} ) \in \overline{V}^{+}_{n}$ if, employing the
    standard coordinates on $\mathscr{C}_{p}$, for all $i = 1$, \dots,
    $n$, $( \zeta_{i} )_{V} > 0$ or $\zeta_{i}$ vanishes.  The
    subscript $V$ here refers to the component along the $V$-direction
    on $\mathscr{C}_{p}$.  Analogously, we say $( x , \zeta ) \in
    \overline{V}^{-}_{n}$ if every $( \zeta_{i} )_{V} < 0$ or
    $\zeta_{i}$ vanishes.  Furthermore, $( x , \zeta ) \in S_{n}$ if
    there exists an index $i$ such that, simultaneously, $\zeta_{i}
    \neq 0$ and $( \zeta_{i} )_{V} = 0$.
  \item \textbf{Smoothness Condition}: The distribution
    $F^{\prime}_{n}$ can be factorised into the tensor product of a
    smooth function and an element of $\mathcal{A}^{n-1}$ when
    localised in a neighbourhood of $V = 0$, \textit{i.e.}, there
    exists a compact set $\mathcal{O} \subset \mathscr{C}_{p}$ such
    that, if $\Theta \in C^{\infty}_{0} ( \mathscr{C}_{p} )$ so that
    it is equal to $1$ on $\mathcal{O}$ and $\Theta^{\prime} \doteq 1
    - \Theta$, then for every multi-index $P$ in $\{ 1 , \dotsc , n
    \}$ and for every $i \leqslant n$,
    \begin{equation}
      \label{eq:factorisation}
      f \doteq \tilde{F}^{\prime}_{n} ( u_{x_{P_{i+1}} , \dotsc ,
        x_{P_{n}}} ) \thinspace \Theta^{\prime}_{x_{P_{1}}}
      \negthinspace \dotsm \thinspace \Theta^{\prime}_{x_{P_{i}}} \in
      C^{\infty} ( \mathscr{C}_{P}^{i} ) \text{,}
    \end{equation}
    where $\tilde{F}^{\prime}_{n} : C^{\infty}_{0} (
    \mathscr{C}_{p}^{n-i-1}) \to \mathcal{D}^{\prime} (
    \mathscr{C}_{p}^{i} )$ is the unique map from $C^{\infty}_{0} (
    \mathscr{C}_{p}^{n-i-1} )$ to $\mathcal{D}^{\prime} (
    \mathscr{C}_{p}^{i} )$ determined by $F_{n}^{\prime}$ using the
    Schwartz kernel theorem.  Furthermore, $u_{x_{P_{i+1}} , \dotsc ,
      x_{P_{n}}} \in C^{\infty}_{0} ( \mathscr{C}^{n-i}_{p} )$, and we
    have specified the integrated variables $x_{P_{i+1}}$, \dots,
    $x_{P_{n}}$.  For every $j \leqslant i$, $\partial_{V_{1}}
    \dotsm \partial_{V_{j}} f$ lies in $C^{\infty} (
    \mathscr{C}_{p}^{i} ) \cap L^{2} ( \mathscr{C}_{p}^{i} ,
    dV_{P_{1}} \wedge d\mathbb{S}_{P_{1}}^{2} \dotsm dV_{P_{i}} \wedge
    d\mathbb{S}_{P_{i}}^{2} ) \cap L^{\infty} ( \mathscr{C}_{p}^{i}
    )$, while the limit of $f$ as $V_{j}$ tends uniformly to $0$ vanishes uniformly in the other coordinates.
  \end{enumerate}
\end{definition}

\renewcommand{\labelenumi}{\alph{enumi})}
\begin{remark}
  \begin{enumerate}
  \item In property~2 of Definition~\ref{Def:testdistributions} we
    required that $\WF ( F_{n}^{\prime} ) \cap S_{n} = \emptyset$,
    \emph{viz.}, no spatial directions are present in the wave front
    set of $F^{\prime}_{n}$.  Even if such an extra condition is not
    stipulated in the definition of the elements on $\mathscr{F}_{e}$,
    here we have to add it because, later on, we want to multiply
    elements of $\mathcal{A}$ with $\omega$ and in the wave front set
    of $\omega$ there are spatial directions, to be specific, the
    intersection $\WF ( \omega ) \cap S_{2}$ is not empty.
  \item Thanks to the smoothness condition, the last requirement in
    Definition~\ref{Def:testdistributions}, the distributions in
    $\mathcal{A}^{n}$ can be extended over $\otimes^{n} \mathscr{S} (
    \mathscr{C}_{p} )$ and such an extension is unique.
  \end{enumerate}
\end{remark}

Notice that $\mathscr{S} ( \mathscr{C}_{p} )$ is strictly contained in
$\mathcal{A}$ and that the candidate to play the role of the extended
algebra on $\mathscr{C}_{p}$ thus is
\begin{equation*}
  \mathscr{A}_{e} ( \mathscr{C}_{p} ) = \bigoplus_{n \geqslant 0}
  \mathcal{A}_{s}^{n} \text{,}
\end{equation*}
where $\mathcal{A}^{n}_{s}$ is the subset of totally symmetric
elements in $\mathcal{A}^{n}$ defined in
Definition~\ref{Def:testdistributions}.  Moreover, the first space in
the previous direct sum is $\mathbb{C}$ and only sequences with a
finite number of elements are considered.  We can now endow this set
with the structure of $^{\ast}$-algebra by introducing the
$^{\ast}$-operation $\{ F^{\prime}_{n} \}^{\ast} \doteq \{
\overline{F^{\prime}_{n}} \}$ for all $F^{\prime} \in
\mathscr{A}_{e}$.  The composition law arises from a modification of
$\star_{B}$ by means of the state constructed in the sequel of
\eqref{eq:state}.  It is \textit{a priori} clear that such a procedure
intrinsically depends on the particular $\omega$ considered.
Nonetheless, our choice will later be justified both through its
connection with the bulk data and by well-posedness of the new
structure.  If we stick to the functional representation, we can thus
introduce
\begin{equation}
  \label{eq:starproductbound}
  \begin{split}
    & \star_{\omega} : \mathscr{A}_{e} ( \mathscr{C}_{p}) \times
    \mathscr{A}_{e} ( \mathscr{C}_{p} ) \to \mathscr{A}_{e} (
    \mathscr{C}_{p} ) \text{,} \\
    & ( F^{\prime} \star_{\omega} G^{\prime} ) ( \Phi ) =
    \sum_{n=0}^{\infty} \frac{1}{n!} \big\langle F^{\prime (n)} ( \Phi
    ) , \omega^{n} G^{\prime (n)} ( \Phi ) \big\rangle \text{,}
  \end{split}
\end{equation}
for all $F^{\prime}$, $G^{\prime} \in \mathscr{A}_{e} (
\mathscr{C}_{p} )$ and for all $\Phi \in C^{\infty} ( \mathscr{C}_{p}
)$.

\begin{proposition}
  \label{Pro:star-omega-product}
  The operation \eqref{eq:starproductbound} is a well-defined product
  in $\mathscr{A}_{e}$.
\end{proposition}
\begin{proof}
  As a starting point, notice that \eqref{eq:starproductbound} is
  bilinear by construction and that, by definition of $\mathscr{A}_{e}
  ( \mathscr{C}_{p})$, there are only a finite number of non-vanishing
  elements $F^{\prime (n)}$ and $G^{\prime (n)}$.  Accordingly,
  \eqref{eq:starproductbound} consists of finite linear combinations
  of terms that formally look like
  \begin{equation}
    \label{eq:passage} 
    \mathcal{S} \int_{\mathscr{C}_{p}^{2k}} F^{\prime}_{j} ( x_{1} ,
    \dotsc , x_{j} ) \medspace \omega ( x_{1} , y_{1} ) \dotsm \omega
    ( x_{k} , y_{k}) \medspace G^{\prime}_{l} ( y_{1} , \dotsc ,y _{l}
    ) \medspace \prod_{i=1}^{k} d\mu ( x_{i} ) \thinspace d\mu(y_{i})
  \end{equation}
  with $k \leqslant j$ and $k \leqslant l$, while $\mathcal{S}$
  realises symmetrisation in the non-integrated variables and $d\mu$ is the
  measure $dV \wedge d\mathbb{S}^{2} ( \theta , \varphi )$ on
  $\mathscr{C}_{p}$, written here in the usual coordinates.
  Therefore, the proof amounts to showing that it is possible to give
  a rigorous meaning to expressions like \eqref{eq:passage} and that
  the result of such an operation is an element of
  $\mathcal{A}^{j+l-2k}$.

  First, we follow the proof of \cite{Hollands:2001aa}, and, to this
  avail, examine if \eqref{eq:passage} can be seen as the target of $1
  \in C^{\infty} ( \mathscr{C}_{p}^{2k} )$ under the linear map
  determined, with the help of the Schwartz kernel theorem, by the
  distribution resulting from multiplication of the two distributions
  $\omega^{\otimes k} \otimes I^{\otimes (j+l-2k)} \in
  \mathcal{D}^{\prime} ( \mathscr{C}^{j+l} )$ and $F^{\prime}_{j}
  \otimes G^{\prime}_{l} \in \mathcal{A}^{j+l}$, where $I$ denotes the
  identity operator on $\mathcal{A}^{1}$.  Let us start by discussing
  the well-posedness of the multiplication of distributions presented
  above.  This can be checked by examining the structure of the wave
  front set of the single objects to verify that their composition
  never contains the zero section.  The key ingredients for this can
  be readily inferred using Theorem~8.2.9 in
  \cite{Hormander_The-Analysis-I:1990},
  \begin{equation}
    \label{eq:WFfg} 
    \WF ( F^{\prime}_{j} \otimes G^{\prime}_{l} ) \subset \big( W_{j}
    \cup \{ 0 \} \big) \times \big( W_{l} \cup \{ 0 \} \big) \setminus
    \{ 0 \} \text{,}
  \end{equation}
  and
  \begin{equation}
    \label{eq:WFomI} 
    \WF ( \omega^{\otimes k} \otimes I^{j+l-2k} ) \subset \big( A \cup
    B \cup \{ 0 \} \big)^{k} \times \{ 0 \} \setminus \{ 0 \} \text{,}
  \end{equation}
  where, as usual, we have not specified the dimension of the zero
  section in the cotangent space.  Furthermore, $A$, $B$ and $W_{j}$
  are defined in \eqref{eq:WFA}, \eqref{eq:WFB} and \eqref{eq:WFS},
  respectively.  It is now possible to apply Theorem~8.2.10 in
  \cite{Hormander_The-Analysis-I:1990} since the above wave front sets
  never sum up to the zero section.  This is tantamount to realising
  that for every $n$ and $m$, since $A^{n} \subset
  \overline{V}^{+}_{n} \times \overline{V}_{n}^{-}$ and
  $\overline{V}_{n}^{\pm} \cap W_{n} = \emptyset$, $B^{n} \times \{
  \mathbb{R}^{3} \}^{m} \cap W_{2n+m} = \emptyset$ and $A^{n} \cap (
  W_{n} \times W_{n} ) = \emptyset$.  The outcome is that, in
  \eqref{eq:passage}, the pointwise product of $F^{\prime}_{j} \otimes
  G^{\prime}_{l} \in \mathcal{D}^{\prime} ( \mathscr{C}_{p}^{j+l} )$
  with $\omega^{\otimes k} \otimes I^{j+l-2k} \in \mathcal{D}^{\prime}
  ( \mathscr{C}_{p}^{j+l} )$ is still a well-defined element of
  $\mathcal{D}^{\prime} ( \mathscr{C}_{p}^{j+l} )$, whose wave front
  set satisfies the inclusion
  \begin{equation}
    \label{eq:WFd}
    \WF \big( F^{\prime}_{j} \otimes G^{\prime}_{l} \cdot
    \omega^{\otimes k} \otimes I^{j+l-2k} \big) \subset \WF (F_{j}
    \otimes G_{l} ) \cup \{ 0 \} + \WF \big( \omega^{\otimes k}
    \otimes I^{j+l-2k} \big) \cup \{ 0 \} \text{,}
  \end{equation}
  where, as usual, the sum of two wave front sets is defined as the
  sum on the fibres of the cotangent spaces.  Unfortunately, this does
  not suffice to show the well-posedness of \eqref{eq:passage}.  Since
  $F^{\prime}_{j}$ and $G^{\prime}_{l}$ are not compactly supported on
  $\mathscr{C}_{p}^{j}$ and $\mathscr{C}_{p}^{l}$, respectively, their
  product does not lie in $\mathcal{E}^{\prime} \big(
  \mathscr{C}_{p}^{j+l} \big)$.  Hence we cannot directly test the
  linear map stemming from $( F^{\prime}_{j} \otimes G^{\prime}_{l} )
  \cdot \big( \omega^{\otimes k} \otimes I^{j+l-2k} \big)$ on the unit
  constant function on $\mathscr{C}_{p}^{2k}$ in order to infer that
  \eqref{eq:passage} is an element in $\mathcal{A}^{j+l-2k}$.

  Let us hence proceed by showing that in the case $k = 1$ we can test
  the linear map arising from $( F^{\prime}_{j} \otimes
  G^{\prime}_{l}) \cdot \big( \omega \otimes I^{j+l-2} \big)$ on $1$
  and that the result of this operation is an element of
  $\mathcal{A}^{j+l-2}$.  The case for a generic $k$ arises from of a
  recursive application of the very same procedure and, eventually,
  the application of an operator realising the total symmetrisation.
  Thus we are interested in
  \begin{equation*}
    \int_{\mathscr{C}_{p}^{2}} ( F^{\prime}_{j} \otimes
    G^{\prime}_{l}) \cdot \big( \omega \otimes I^{j+l-2} \big)
    \medspace d\mu ( x_{1} ) \thinspace d\mu ( y_{1} ) \text{,}
  \end{equation*}
  where $F^{\prime}_{j} \in \mathcal{A}^{j}$ and $G^{\prime}_{l} \in
  \mathcal{A}^{l}$.  We exploit property~$3$ of
  Definition~\ref{Def:testdistributions} and notice that, if the
  smoothness condition holds for a compact set $\mathcal{O}$, it also
  holds for every larger compact set $\mathcal{O}_{1}$ containing
  $\mathcal{O} \in \mathscr{C}_{p}$.  We can thus find a common set
  $\mathcal{O}_{1}$ for which the smoothness condition property is
  true at the same time for $F^{\prime}_{j} = ( \Theta +
  \Theta^{\prime} ) \thinspace F^{\prime}_{j}$ and $G^{\prime}_{l} = (
  \Theta + \Theta^{\prime} ) \thinspace G^{\prime}_{l}$ with respect to
  a common compactly supported function $\Theta$ equal to $1$ on
  $\mathcal{O}_{1}$.  Effectively, the above integral is divided into
  the sum of four different ones, which we now analyse separately.

  \textit{Part I)} The first term is
  \begin{equation}
    \label{eq:integrationoneomega}
    \int_{\mathscr{C}^{2}_{p}} ( \Theta ( x_{1} ) F^{\prime}_{j}
    \otimes \Theta ( y_{1} ) G^{\prime}_{l} ) \cdot \big( \omega
    \otimes I^{j+l-2} \big ) \medspace d\mu ( x_{1} ) \thinspace d\mu
    ( y_{1} ) \text{.}
  \end{equation}
  In this case the integral can be considered as the smearing of a
  distribution in $\mathcal{D}^{\prime} \big( \mathscr{C}^{j+l}_{p}
  \big)$ with a test function in $C^{\infty}_{0} ( \mathscr{C}^{2}_{p}
  )$.  Hence, Theorem~8.2.12 of \cite{Hormander_The-Analysis-I:1990}
  ensures that, using the notation introduced there, the result of
  \eqref{eq:integrationoneomega} is a distribution whose wave front
  set is contained in $W_{j+l-2} \cup ( W_j \times W_l ) \circ ( A
  \times \{ 0 \} ) \subset W_{j+l-2}$ as given in \eqref{eq:WFS}.
  Notice that, in the proof of the last inclusion, we have used
  \eqref{eq:WFfg} and \eqref{eq:WFomI}.  Hence property~$2$ in
  Definition~\ref{Def:testdistributions} holds.  That said,
  property~$1$ is automatically satisfied since, by hypothesis,
  $F^{\prime}_{j} \in \mathcal{A}^{j}$ and $G^{\prime}_{l} \in
  \mathcal{A}^{l}$, while property~$3$ holds true for the resulting
  distribution as \eqref{eq:factorisation} is valid \textit{a priori}
  in all variables and, thus, left untouched for those which have not
  been integrated out in \eqref{eq:integrationoneomega}.

  \textit{Part II)} The second term is
  \begin{equation}
    \label{eq:secondterm}
    \int_{\mathscr{C}^{2}_{p}} ( \Theta^{\prime} ( x_{1} )
    F^{\prime}_{j} \otimes \Theta^{\prime} ( y_{1} )
    G^{\prime}_{l} ) \cdot \big(\omega \otimes I^{j+l-2} \big )
    \medspace d\mu ( x_{1} ) \thinspace d\mu ( y_{1} ) \text{.}
  \end{equation}
  The analysis is rather simple if we make profitable use of
  \eqref{eq:factorisation} in the integrated variables and interpret
  the previous integral in the weak sense.  Namely, let $f \doteq
  \Theta^{\prime} ( x_{1} ) F^{\prime}_{j} ( u )$ for some $u \in
  C^{\infty}_{0} \big( \mathscr{C}_{p}^{j-1} \big)$ and $f^{\prime}
  \doteq \Theta^{\prime} ( x_{j+1} ) G^{\prime}_{l} ( u^{\prime} )$
  for some $u^{\prime} \in C^{\infty}_{0} \big( \mathscr{C}_{p}^{l-1}
  \big)$, then we have that the operation $\omega ( f , f^{\prime} )$
  is well defined due to the continuity property
  \eqref{eq:2pcontinuity} satisfied by $\omega$.  Hence, properties
  $1$, $2$ and $3$ in Definition~\ref{Def:testdistributions} hold true
  because they are satisfied by $\Theta^{\prime} ( x_{1} )
  F^{\prime}_{j} \otimes \Theta^{\prime} ( y_{1} ) G^{\prime}_{l}$.

  \textit{Parts III \& IV)} The remaining two terms are substantially
  identical and we treat only one of them.  Hence consider
  \begin{equation}
    \label{eq:termIII}
    \int_{\mathscr{C}^{2}_{p}} ( \Theta F^{\prime}_{j} \otimes
    \Theta^{\prime} G^{\prime}_{l} ) \cdot \big( \omega \otimes
    I^{j+l-2} \big) \medspace d\mu ( x_{1} ) \thinspace d\mu ( y_{1} )
    \text{.}
  \end{equation}
  In order to cope with this integral we introduce a new larger
  factorisation $\eta + \eta^{\prime} = 1$ with $\eta \in
  C^{\infty}_{0} ( \mathscr{C}_{p} )$ such that $\eta = 1$ on a large
  compact set properly containing the closure of $\supp ( \Theta )$ so
  that both $\supp ( \eta^{\prime} \Theta^{\prime} ) \cap \supp (
  \Theta ) = \emptyset$ and $\eta^{\prime} \Theta^{\prime} =
  \eta^{\prime}$.  If now $G^{\prime}_{l}$ is substituted with $( \eta
  + \eta^{\prime} ) G^{\prime}_{l}$ we obtain another splitting.  On
  the one hand, since $\Theta^{\prime} \eta \in C^{\infty}_{0} (
  \mathscr{C}_{p} )$, the analysis of $\Theta F^{\prime}_{j} \otimes
  \Theta^{\prime} \eta G^{\prime}_{l}$ boils down to that of case~I,
  while, on the other hand, $\frac{\Theta ( V ) [ \Theta^{\prime}
    \eta^{\prime} ] ( V^{\prime} )}{( V - V^{\prime} )^{2}}$ turns out
  to be smooth on $\mathscr{C}_{p}$, since by construction $( V -
  V^{\prime} )^{2} > 0$ for $V$ on the support of $\Theta$ and
  $V^{\prime}$ on that of $\eta^{\prime}$.  Hence, if we write the
  smoothness condition \eqref{eq:factorisation} by means of $\eta$ as
  $\eta^{\prime} ( x_{j+1} ) G^{\prime}_{l} ( t_{l-1} ) = f ( x_{j+1}
  )$, where $t_{l-1} \in C^{\infty}_{0} \big( \mathscr{C}_{p}^{l-1}
  \big)$, we obtain that $u \doteq \Theta \omega ( f )$ is a compactly
  supported smooth function on $\mathscr{C}$, thus yielding that
  $F^{\prime}_{j} ( u )$ is a well-posed operation as $u$ is compactly
  supported.  Furthermore, in order to conclude the analysis of the
  present case, due to Theorem~8.2.12 in
  \cite{Hormander_The-Analysis-I:1990}, we notice that the wave front
  set of \eqref{eq:termIII} is contained in $W_{j+l-2}$ given in
  \eqref{eq:WFS} and that property~3 in
  Definition~\ref{Def:testdistributions} holds true just by applying
  \eqref{eq:factorisation} before smearing it.  \qed
\end{proof}

The result of this subsection is that $\big( \mathscr{A}_{e} (
\mathscr{C}_{p} ) , \star_{\omega} \big)$ is a full-fledged
topological $^{\ast}$-algebra.  Furthermore, due to the compactness
property stated in Definition~\ref{Def:testdistributions}, the
subalgebra $( \mathscr{A}_{e} \big( \mathscr{C}_{p}^{+} ) ,
\star_{\omega} \big)$ defined by restriction of the domain of the test
distributions to $\mathscr{C}_{p}^{+}$ is a well-defined topological
$^{\ast}$-algebra and, thus, we are ready to discuss the intertwining
relations between bulk and boundary data.

\subsection{Interplay Between the Algebras and the States on
  $\mathscr{D}$ and on $\mathscr{C}_{p}^{+}$}
\label{ssec:3.5}

We are now in the position to discuss a connection between the field
theories described above, hence setting up a bulk-to-boundary
correspondence and identifying an Hadamard state in the bulk.  The
whole subsection is devoted to this issue, but, as a starting point,
we need to recapitulate the geometric structure in order to clearly
relate Subsections~\ref{ssec:2.2} and \ref{ssec:3.2}.

Recall that we consider the globally hyperbolic subset
$\mathcal{O}^{\prime}$ contained in a geodesic neighbourhood of an
arbitrary but fixed point $p$ in a strongly causal spacetime $M$.  In
$\mathcal{O}^{\prime}$ we single out a double cone $\mathscr{D} \equiv
\mathscr{D} ( p , q )$, which plays the role of the bulk spacetime,
while the set $\partial J^{+} ( p ) \cap \overline{\mathscr{D}}$ is
our selected boundary.  Up to the choice of an orthonormal frame in
$p$, the latter can be seen as the locus $u = 0$ in the natural
coordinate system $\big( u , r , x^{A} \big)$ introduced in
Subsection~\ref{ssec:2.2} which is furthermore endowed with the metric
\eqref{eq:metriccone}.  In terms of the structure of
Subsection~\ref{ssec:3.2}, we can identify the boundary as
$\mathscr{C}^{+}_{p}$ with a small caveat with respect to the
coordinates used.  While it is always possible to switch from $x^{A}$
($A = 1$, $2$) to the standard $( \theta , \varphi )$, the role of $V$
as a coordinate is played by $r$, the affine parameter on the null
geodesics of the cone.  As a last point, the role of the function $h$
in \eqref{eq:mS} is taken in general by $\sqrt[4]{\abs{g_{AB}}}$,
where $g_{AB}$ are the metric components appearing in
\eqref{eq:metriccone} evaluated at $u = 0$ and
$\abs{\medspace\cdot\medspace}$ is kept to indicate the determinant.
It is interesting to notice, that, whenever the conditions for the
reduction of \eqref{eq:metriccone} to \eqref{eq:metriccone2} are
fulfilled, $h$ can be set to $V \equiv r$, (see also the relation
between the volume elements of the sphere in different coordinates
\eqref{eq:spherevolumeelement}).  Furthermore, in the retarded
coordinates used, the exponential map becomes an identity. Hence, if
not strictly necessary, we shall not indicate it anymore.

Now we proceed in two steps. The first one consists in proving the
possibility of introducing a well-defined map from the extended
algebra in $\mathscr{D}$ to the one on $\mathscr{C}^{+}_{p} \subset
\mathscr{C}_{p}$, while, in the second, we prove that this map is also
well-behaved with respect to the algebra structures.
\begin{theorem}
  \label{The:Pi1}
  Let $\mathscr{D}$ be a double cone and regard the portion
  $\mathscr{C}^{+}_{p}$ of the boundary as part of a cone
  $\mathscr{C}_{p}$.  Let us introduce the linear map $\Pi :
  \mathscr{F}_{e} ( \mathscr{D} ) \to \mathscr{A}_{e} (
  \mathscr{C}^{+}_{p} )$ by setting
  \begin{equation}
    \label{eq:restriction}
    \Pi_{n} ( F_{n} ) \doteq \sqrt[4]{\abs{g_{AB}}}_{1} \dotsm
    \sqrt[4]{\abs{g_{AB}}}_{n} \medspace \Delta^{\otimes n} ( F_{n} )
    \big\vert_{(\mathscr{C}_{p}^+)^{n}} \text{,}
  \end{equation}
  where $\Delta$ is the causal propagator \eqref{eq:causalp},
  $\vert_{\mathscr{C}^{+}_{p}}$ denotes the restriction on
  $\mathscr{C}^{+}_{p}$ and the subscripts $1$, \dots ,$n$ entail
  dependence of the root on the coordinates of the $i$-th cone.  Then,
  the following properties hold true:
  \renewcommand{\labelenumi}{\arabic{enumi})}
  \begin{enumerate}
  \item $\hat{\Pi}_{n}$, the integral kernel of $\Pi_{n}$, is equal to
    $\otimes^{n} \hat{\Pi}_{1}$ and is an element of
    $\mathcal{D}^{\prime} \big( ( \mathscr{C}^{+}_{p} \times
    \mathscr{D} )^{n} \big)$.  The wave front set of $\hat{\Pi}_{n}$
    satisfies
    \begin{equation}
      \label{WFPin}
      \WF ( \hat{\Pi}_{n} ) \subset ( \WF ( \hat{\Pi}_{1} ) \cup \{ 0
      \} )^{n} \setminus \{ 0 \} \text{.}
    \end{equation}
    Furthermore, if $( x , \zeta_{x} ; y , \zeta_{y} ) \in \WF (
    \hat{\Pi}_{1} )$, then:
  {\renewcommand{\labelenumi}{(\alph{enumi})}
    \begin{enumerate}
    \item neither $\zeta_{x}$ nor $\zeta_{y}$ vanish;
    \item $( \zeta_{x} )_{r} \neq 0$
    \item $( \zeta_{x} )_{r} \geqslant 0$ if and only if $-\zeta_{y}$
      is future directed.
    \end{enumerate}}
  \item The image of $\mathscr{F}_{e} ( \mathscr{D} )$ under $\Pi$
    lies in $\mathscr{A}_{e} ( \mathscr{C}_{p})$.
  \end{enumerate}
\end{theorem}
\begin{proof}
  We prove the above properties in two separate steps.

\vskip .2cm

\noindent \textit{I) Construction of $\big( \sqrt[4]{\abs{g_{AB}}}
      \medspace \Delta_{\mathscr{C}} \big)^{\otimes n}$ and the wave
      front set of its integral kernel}

    In the normal neighborhood $\mathcal{O}_{p} \subset M$ which
    contains $\mathscr{D}$ we can select a subset
    $\mathcal{O}^{\prime} \subset ( J^{-} ( \mathscr{D} ) \setminus
    J^{-} ( p ) ) \subset \mathcal{O}_{p} $ which is a globally
    hyperbolic open set that extends $\mathscr{D}$ over
    $\mathscr{C}_{p}^{+}$, but neither over $p$ nor over the future of
    $\mathscr{D}$.  The existence of a similar set results from the
    global hyperbolicity of $M$ which contains $\mathcal{O}_{p}$ and
    thus also $\mathscr{D}$.  Let us indicate by $\hat{\Delta} \in
    \mathcal{D}^{\prime} ( \mathcal{O}^{\prime} \times \mathscr{D} )$
    the integral kernel of $\Delta : C^{\infty}_{0} ( \mathscr{D} )
    \to C^{\infty} ( \mathcal{O}^{\prime} )$ defined by restricting
    the map in \eqref{eq:causalp}.  It holds true that
    \begin{equation}
      \label{eq:WFDe} 
      \WF ( \hat{\Delta} ) = \big\{ ( x_{1} , \zeta_{1} ; x_{2} ,
      \zeta_{2} ) \in T^{\ast} \mathcal{O}^{\prime} \times T^{\ast}
      \mathscr{D} \setminus \{ 0 \} \medspace \big\vert \medspace (
      x_{1} , \zeta_{1} ) \sim ( x_{2} , -\zeta_{2} ) \big\} \text{,}
    \end{equation}
    where the equivalence relation $( x_{1} , \zeta_{1} ) \sim ( x_{2}
    , \zeta_{2} )$ means that there exists a null geodesic $\gamma$
    with respect to the metric $g$ in $\mathscr{D}$ which contains
    both $x$ and $y$.  Furthermore, $g^{\mu\nu} ( \zeta_{1} )_{\nu}$
    and $g^{\mu\nu} ( \zeta_{2} )_{\nu}$ are the tangent vectors of
    the affinely parametrised geodesic $\gamma$ in $x$ and $y$,
    respectively \cite{Radzikowski:1996aa,Radzikowski:1996ab}.

    We proceed by restricting one entry of the causal
    propagator\footnote{The same procedure was employed in the proof
      of Proposition~4.3 in \cite{Moretti:2008aa} or in the work
      \cite{hollands:2000}.} on $\mathscr{C}_{p}^{+}$, while leaving
    the other localised in $\mathscr{D}$.  To this end, let us define
    the embedding $\chi$ of $\mathscr{C}_{p}^{+} \times \mathscr{D}$
    to $\mathcal{O}^{\prime} \times \mathscr{D}$, whose action, in
    retarded coordinates on $\mathcal{O}^{\prime}$, is defined as
    $\chi (r , \theta , \varphi ; x_{2} ) = ( 0 , r , \theta , \varphi
    ; x_{2} )$.  According to Theorem~8.2.4 of
    \cite{Hormander_The-Analysis-I:1990}, the restriction of the first
    entry of $\hat{\Delta}$ on $\mathscr{C}_{p}^{+}$ by means of the
    pullback under $\chi$ is well defined provided that     
    $M_{\chi} \cap
    \WF ( \hat{\Delta} ) = \emptyset$, where $M_{\chi}$ is the set of
    normals of $\chi$.  In order to verify this statement about the
    empty intersection, we notice that, using the definition employed
    in such a theorem,
    \begin{equation*}
      M_{\chi}\subset N_{\chi} = \big\{ ( x_{1} , \zeta_{1} ; x_{2} , \zeta_{2} ) \in
      T^{\ast} \mathcal{O}^{\prime} \times T^{\ast} \mathscr{D}
      \medspace \big\vert \medspace x_{1} \in \mathscr{C}_{p}^{+}\subset\mathcal{O}^\prime
      \text{,~} \zeta_{1} = ( \zeta_{1} )_{u} du \text{,~}(\zeta_{1})_u \in \mathbb{R} \big\} \text{.}
    \end{equation*}
   We shall now prove that $N_{\chi} \cap
    \WF ( \hat{\Delta} ) = \emptyset$, 
    consider $( x_{1} , \zeta_{1} ; x_{2} , \zeta_{2} ) \in N_{\chi}$
    and the null geodesic $\gamma^{\prime}$ originating from $x_{1}$
    whose tangent vector in $x_{1}$ is equal to $g^{-1} ( \zeta_{1}
   )$.  Notice that, in the retarded coordinates, the only
    non-vanishing component of $g^{-1} ( \zeta_{1} )$ is $( g^{-1} (
    \zeta_{1} ) )^{r}$ which implies that $\gamma^{\prime}$ is
    contained in $\mathscr{C}_{p}^{+}$ and, in particular, does not
    enter $\mathscr{D}$.  For this reason the intersection of
    $N_{\chi}$ with $\WF ( \hat{\Delta} )$ is the empty set and hence
    the hypotheses of Theorem~8.2.4 of
    \cite{Hormander_The-Analysis-I:1990} are fulfilled.  Thus
    $\hat{\Delta}_{\mathscr{C}}$, the pullback of $\hat{\Delta}$ under
    $\chi$, can be defined in one and only one way and $\WF (
    \hat{\Delta}_{\mathscr{C}} ) \subset \chi^{\ast} \WF (
    \hat{\Delta} )$.  In particular, this entails that, if $( x ,
    \zeta_{x} ; y , \zeta_{y} ) \in \WF ( \hat{\Delta}_{\mathscr{C}}
    )$ with $x \in \mathscr{C}_{p}^{+}$ and $y \in \mathscr{D}$, it
    enjoys properties (a), (b) and (c) stated above.

    In order to verify (b), suppose this were not true and consider $(
    x , \zeta_{x} ; y , \zeta_{y} ) \in \WF (
    \hat{\Delta}_{\mathscr{C}} )$, where $( \zeta_{x} )_{r} = 0$.
    Thus there should exist an element $( x , \zeta^{\prime}_{x} ; y ,
    \zeta_{y} )$ such that $\chi^* ( x , \zeta^{\prime}_{x} ; y ,
    \zeta_{y} ) = ( x , \zeta_{x} ; y , \zeta_{y} )$, where
    $\zeta^{\prime}_{x}$ is a null covector whose components are $(
    \zeta^{\prime}_{x} )_{r} = 0$, $( \zeta^{\prime}_{x} )_{\theta} =
    ( \zeta_{x} )_{\theta}$ and $( \zeta^{\prime}_{x} )_{\varphi} = (
    \zeta_{x} )_{\varphi}$ while $( \zeta^{\prime}_{x} )_{u}$ is a
    fixed number in $\mathbb{R}$.  Since $g^{-1} (
    \zeta^{\prime}_{x})$ has to be null and since $( g^{-1} (
    \zeta^{\prime}_{x} ) )^{u} = 0$, the only possibility is $(
    \zeta^{\prime}_{x} )_{\theta} = ( \zeta^{\prime}_{x} )_{\varphi} =
    0$.  Hence the only non-vanishing component of $g^{-1} (
    \zeta^{\prime}_{x} )$ is the $r$-component, which implies that
    $\zeta^{\prime}_{x} = ( \zeta^{\prime}_{x} )_{u} du$, thus $( x ,
    \zeta^{\prime}_{x} ; y , \zeta_{y} ) \in N_{\chi}$.  At this point
    we have reached a contradiction because $\WF ( \hat{\Delta} ) \cap
    N_{\chi} = \emptyset$ so that $( \zeta_{x} )_{r}$ has to be
    different from zero.  Notice that (a) and (c) result from the
    constraint imposed by the equivalence relation $\sim$ in the wave
    front set of $\hat{\Delta}$ in \eqref{eq:WFDe} and from the
    observation that the projection of $\zeta^{\prime}_{x}$ on
    $\mathscr{C}_{p}^{+}$ under $\chi^{\ast}$ does not change the
    causal direction (past/future).  In order to accomplish this part
    of the proof, we only need to multiply every
    $\hat{\Delta}_{\mathscr{C}}$ with $\sqrt[4]{\abs{g_{AB}}} \otimes
    1$ which is smooth because it is the $4$-th order root of a smooth
    positive function.  Hence the wave front set of the resulting
    distribution is left unchanged.  Thus we define $\hat{\Pi}_{n}$ as
    the tensor product of distributions $\hat{\Pi}_{n} \doteq ( h
    \thinspace \hat{\Delta}_{\mathscr{C}} )^{\otimes n} \in
    \mathcal{D}^{\prime} ( ( \mathscr{C}_{p}^{+} \times \mathscr{D}
    )^{n} )$, where $h = \sqrt[4]{\abs{g_{AB}}} \otimes 1$, and, due
    to Theorem~8.2.9 in \cite{Hormander_The-Analysis-I:1990}, $\WF (
    \hat{\Pi}_{n} )$ enjoys the inclusion \eqref{WFPin}.  By the
    Schwartz kernel theorem we obtain the linear map $\Pi_{n} = (
    \sqrt[4]{\abs{g_{AB}}} \thinspace \Delta_{\mathscr{C}} )^{\otimes
      n}$ whose integral kernel is $\hat{\Pi}_{n}$.

\vskip .2cm

\textit{II) On the image of $\Pi$}

    Notice that, since every $F \in \mathscr{F}_{e} ( \mathscr{D} )$
    is composed of a finite number of $F_{n}$, it is sufficient to
    prove that the generic $F_{n}$ is mapped to an element of
    $\mathcal{A}_{s}^{n}$ by $\Pi$ or, rather, by $\Pi_{n}$.
    Moreover, the pointwise product of $\hat{\Pi}_{n}$, the integral
    kernel of $\Pi_{n}$, and $I^{n} \otimes F_{n}$, $I$ the unit
    constant function in $\mathcal{D}^{\prime} ( \mathscr{C}_{p} )$,
    is well-defined because their wave front sets do not sum up to the
    zero section, as one can infer from \eqref{eq:inters}, from
    Theorem~8.2.9 in \cite{Hormander_The-Analysis-I:1990} and from
    \eqref{WFPin} along with the discussion presented above.  More
    precisely, we have that $\WF ( I^{n} \otimes F_{n} ) \subset \{ 0
    \} \times \WF ( F_{n} )$ where $\{ 0 \}$ is the zero section in
    $T^{\ast} {\mathscr{C}_{p}^{+}}^{n}$ while every element in $\WF (
    \hat{\Pi}_{n})$ has non-vanishing components on that cotangent
    space, thus $\WF ( I^{n} \otimes F_{n} ) + \WF ( \hat{\Pi}_{n})$
    cannot contain the zero section.  Hence the H\"ormander criterion
    for the multiplication of distributions, Theorem~8.2.10 in
    \cite{Hormander_The-Analysis-I:1990}, is fulfilled and, moreover,
    the resulting distribution $( \hat{\Pi}_{n} ) \cdot ( I^{n}
    \otimes F_{n} )$ can be tested on any compactly supported smooth
    characteristic function $\eta$ on the support of $F_{n}$ yielding
    the $\Pi_{n} ( F_{n} )$ sought-after. Theorem~8.2.12 in
    \cite{Hormander_The-Analysis-I:1990} guarantees that $\Pi_{n} (
    F_{n} ) \in \mathcal{D}^{\prime} \big( {\mathscr{C}_{p}^{+}}^{n}
    \big)$ and that $\WF ( \Pi_{n} ( F_{n} ) ) \subset \big\{ (x_{1} ,
    \zeta_{x_1} ; \dotsc ; x_{n} , \zeta_{x_n} ; y_{1} , 0 ; \dotsc ;
    y_{n} , 0 ) \in \WF \big( ( \hat{\Pi}_{n} ) \cdot ( I^{n} \otimes
    F_{n} ) \big) \big\} \subset W_{n}$ as in \eqref{eq:WFS}.  In
    order to verify the last inclusion, we notice that $\WF (
    \hat{\Pi}_{1} ) \cap ( S_{1} \times T^{\ast} \mathscr{D} ) =
    \emptyset$ and that, making once more use of Theorem~8.2.10 in
    \cite{Hormander_The-Analysis-I:1990}, $WF( \hat{\Pi}_{n} ) \cdot (
    I^{n} \otimes F_{n} ) \cap ( \overline{V}_{n}^{\pm} \times \{ 0 \}
    ) = \emptyset$, where $\{ 0 \}$ is the zero section in $T^{\ast}
    \mathscr{D}^{n}$.  Thus the wave front set of $\Pi_{n} ( F_{n} )$
    is contained in $W_{n}$, and this is tantamount to the second
    condition in Definition \ref{Def:testdistributions}.

    As for the first one, this can be shown to hold true since, by
    construction, $\supp ( F_{n} ) \subset K^{n} \subset
    \mathscr{D}^{n}$, $K$ a compact set, and hence, due to the support
    property of $\Delta$, there exists another compact set
    $K^{\prime}$ constructed as the closure of $J^{-} ( K ) \cap
    \mathscr{C}_{p}^{+}$ in $\mathscr{C}$ such that ${K^{\prime}}^{n}$
    contains the support of $\Pi_{n} ( F_{n} )$.

    The third and last requirement can also be established by
    recalling that the singular support of the causal propagator
    \eqref{eq:causalp} is contained in the set of the null geodesics.
    Furthermore, those emanating from the support of any $F_{n}$
    (recall that $\supp ( F_{n} ) \subset K$) intersect
    $\mathscr{C}_{p}$ on a compact set that is disjoint from $p$ in
    particular.  Hence the causal propagator is a smooth function
    whenever one entry is smoothly localised\footnote{The localisation
      is realised by pointwise multiplication with smooth functions of
      suitable support.} on the support of $F_{n}$ and the other one
    on a neighbourhood of $p$.  Furthermore, even after multiplication
    both by the function $\Theta^{\prime}$ as in
    Definition~\ref{Def:testdistributions} and by
    $\sqrt[4]{\abs{g_{AB}}}$, such smooth function is
    square-integrable and bounded, together with its $V$-derivative,
    in a suitable open set of $\mathscr{C}_{p} \sim \mathbb{R}^{+}
    \times \mathbb{S}^{2}$ such that $V \in ( 0 , V_{0} )$, because it
    is a restriction to $\mathscr{C}^{+}_{p}$ of a smooth function
    defined in a neighbourhood of $p$ multiplied by
    $\sqrt[4]{\abs{g_{AB}}}$.  For the same reason, the limit $V \to
    0$ of $\sqrt[4]{\abs{g_{AB}}} \thinspace
    \hat{\Delta}_{\mathscr{C}}$ tends to zero whenever one entry of
    the causal propagator is localised on some compact set in
    $\mathscr{D}$.  Finally, notice that $\Pi_{n} ( F_{n} )$ is
    totally symmetric whenever $F_{n}$ has this property, and this
    completes the proof that $\Pi_{n} ( F_{n} ) \in
    \mathcal{A}^{n}_{s}$.
\end{proof}

As an intermediate step, we proceed by discussing the effect of the
map $\Pi$ on the symplectic form.
\begin{proposition}
  \label{Pro:thesymplecticformisconserved}
  The projection $\Pi : \mathscr{F}_{b} ( \mathscr{D} ) \to
  \mathscr{A}_{e} ( \mathscr{C}_{p}^{+} )$ is a symplectomorphism,
  \textit{i.e.}, for every $f$, $h \in C^{\infty}_{0} ( \mathscr{D}
  )$,
  \begin{equation}
    \label{eq:symplcons}
    \sigma ( \varphi_{f} , \varphi_{h} ) = \sigma_{\mathscr{C}} (
    \Pi_{1} f , \Pi_{1} h ) \text{,}
  \end{equation}
  with $\sigma$ taken as in \eqref{eq:symplbulk}.
\end{proposition}
\begin{proof}
  Let $\varphi_{f} = \Delta f$ and $\varphi_{h} = \Delta h$, where
  $\Delta$ is as in \eqref{eq:causalp}, and consider both a Cauchy
  surface $\Sigma$ of $\mathscr{D}$ and the portion of
  $\mathcal{O}_{1} \doteq \mathscr{D} \cap I^{-} ( \Sigma )$ whose
  boundary is formed by the null surface $\mathscr{C}^{+}_{p}$ and by
  $\Sigma$.  Then the current
  \begin{equation*}
    J_{\mu} \doteq \varphi_{f} \partial_{\mu} \varphi_{h} -
    \varphi_{h} \partial_{\mu} \varphi_{f}
  \end{equation*}
  satisfies $\int_{\Sigma} d\mu ( \Sigma ) \medspace n^{\mu} J_{\mu} =
  \sigma ( \varphi_{f} , \varphi_{h})$ with $n^{\mu}$ the unit future
  directed vector normal to $\Sigma$.  Hence we can apply the
  divergence theorem to $J_{\mu}$ in $\mathcal{O}_{1}$ considered as a
  subregion of a larger globally hyperbolic spacetime,
  $\mathcal{O}^{\prime}$, that contains $\mathscr{D}$.  The result is
  that, since $\nabla^{\mu} J_{\mu} = 0$ in $\mathcal{O}_{1}$ in
  particular, the following identity holds,
  \begin{equation*}
    \sigma ( \varphi_{f} , \varphi_{h} ) = \int_{\mathscr{C}^{+}_{p}}
    d\mu ( \mathscr{C}^{+}_{p} ) \medspace n^{\mu} J_{\mu} \text{.}
  \end{equation*}
  Furthermore, the right-hand side of the preceding equation can be
  rewritten in terms of the retarded coordinates on $\mathscr{C}_{p}^{+}$
  and, if one uses the relation between the volume elements on the
  sphere \eqref{eq:spherevolumeelement} in spherical and local
  coordinates, it becomes
  \begin{equation}
    \label{eq:symplaux}
    \int_{\mathbb{R}^{+} \negthinspace \times \mathbb{S}^{2}}
    \sqrt{\abs{g_{AB}}} \Big[ \varphi_{f} \frac{\partial}{\partial r}
    \varphi_{h} - \varphi_{h} \frac{\partial}{\partial r} \varphi_{f}
    \Big] \thinspace dr \wedge d\mathbb{S}^{2} \text{,}
  \end{equation}
  where both $\varphi_{f}$ and $\varphi_{h}$ are evaluated on
  $\mathscr{C}^{+}_{p}$, a legitimate operation as explained at the
  beginning of this section, and, furthermore, they vanish on the
  complement of $\mathscr{C}^{+}_{p}$ in $\mathscr{C}_{p}$.  Finally,
  due to the antisymmetry of the preceding expression, we can consider
  $\sqrt[4]{\abs{g_{AB}}} \thinspace \varphi_{f} = \Pi_{1} f$ as well
  as $\sqrt[4]{\abs{g_{AB}}} \thinspace \varphi_{h} = \Pi_{1} h$, and
  a direct inspection shows that \eqref{eq:symplaux} equals
  $\sigma_{\mathscr{C}} ( \Pi_{1} f , \Pi_{1} g )$ as given in
  \eqref{eq:symplecticform} setting $r = V$. \qed
\end{proof}

\begin{remark}
  Notice that the ideal $\mathscr{I}$ generated by the equations of
  motion \eqref{eq:eqm} is mapped by $\Pi$ to $0 \in \mathscr{A}_{e} (
  \mathscr{C} )$, because $\Delta$ is a weak solution of
  \eqref{eq:eqm}.  Hence the image of both $\mathscr{F}_{b} (
  \mathscr{D} )$ and $\mathscr{F}_{bo} ( \mathscr{D} )$ under $\Pi$
  lie in $\mathscr{A}_{e} ( \mathscr{C} )$; actually they coincide.
\end{remark}

On the basis of this remark, we stress another important property of
$\Pi$.
\begin{proposition}
  \label{Pro:injectivity}
  The map $\Pi$ is injective when acting on the on-shell extended
  algebra $\mathscr{F}_{eo} ( \mathscr{D} )$.
\end{proposition}
\begin{proof}
  Let us recall that the action of $\Pi$ on $F = \{ F_{n} \}_{n} \in
  \mathscr{F}_{e} ( \mathscr{D})$ is determined by the actions of
  $\Pi_{n} = \Pi_{1}^{\otimes n}$ on its components $F_{n}$.  We shall
  hence analyse the kernel of $\Pi_{1}$ seen as a map from
  $\mathcal{T}^{\prime} ( \mathscr{D} )$ to some functionals on
  $\mathscr{S} ( \mathscr{C}^{+}_{p} )$, where the elements of
  $\mathcal{T}^{\prime} ( \mathscr{D} )$ are the compactly supported
  symmetric distributions whose wave front sets enjoy
  \eqref{eq:inters}.  We shall prove that $\ker ( \Pi_{1} ) =
  \mathscr{K} \doteq \big\{ P u \thinspace \big\vert \thinspace u \in
  \mathcal{T}^{\prime} ( \mathscr{D} ) \big\}$.  Given $u \in
  \mathcal{E}^{\prime} ( \mathscr{D} )$ there exists a sequence of
  $u_{j} \in C^{\infty}_{0} ( \mathscr{D} )$ whose support is
  contained in $K$, a proper compact subset of $\mathscr{D}$, such
  that $u_{j} \to u$ weakly for $j \to \infty$.  Consider then the
  following chain of equalities
  \begin{equation*}
    -\int_{M} \Delta ( f ) \thinspace u_{j} = \int_{M} \Delta ( f )
    \thinspace P \Delta_{A} ( u_{j} ) = \sigma ( \Delta ( f ) , \Delta
    ( u_{j} ) ) = \sigma_{\mathscr{C}} ( \Pi_{1} ( f ) , \Pi_{1} (
    u_{j} ) ) \text{,}
  \end{equation*}
  where $P$ is the operator realising the equation of motion and
  $\Delta_{A}$ is its advanced fundamental solution.  Furthermore,
  $\Sigma$ does intersect neither $K$ nor $J^{+} (K)$, and hence, on
  $\Sigma$, $\Delta_{A} ( u_{j} )$ is equal to $\Delta_{R} ( u_{j} )$.
  Notice that in order to obtain the second equality, we have to
  choose two elements of a family of Cauchy hypersurfaces, $\Sigma$
  and $\Sigma^{\prime}$, which do not intersect $K$ and such that
  $J^{+} ( K ) \cap \Sigma = \emptyset$ and $J^{-} ( K ) \cap
  \Sigma^{\prime} = \emptyset$, and integrate by parts twice.  The
  resulting integral on $\Sigma^{\prime}$ vanishes due to the support
  property of $\Delta_{A}$, while the integral on $M$ is zero since $P
  ( \Delta ( f ) ) = 0$.  Thus the remaining integral is precisely the
  symplectic form computed on $\Sigma$.  Furthermore, the last
  equality derives from
  Proposition~\ref{Pro:thesymplecticformisconserved}.  We proceed by
  writing
  \begin{equation*}
    \int_{M} \Delta ( f ) \thinspace u_j = - 2 \int_{\mathscr{C}_{p}}
    \sqrt[4]{\abs{g_{AB}}} \thinspace \Delta ( u_{j} )
    \frac{\partial}{\partial V} \Big( \sqrt[4]{\abs{g_{AB}}}
    \thinspace \Delta ( f ) \Big) dV d\mathbb{S}^{2} \text{.}
  \end{equation*}
  Passing now to the weak limit yields $\Pi_{1} ( u ) (
  -2 \partial_{V} \Pi_{1} f ) = u ( \Delta ( f ) )$.  From the
  previous discussion we obtain that, letting $S \doteq
  -2 \partial_{V} \Pi_{1} ( C^{\infty}_{0} ( \mathscr{D} ) )$, the
  condition $\Pi_{1} ( u ) ( S ) = 0$ is equivalent to $u ( \Delta (
  \mathcal{D} ( \mathscr{D} ) ) ) = 0$, and the latter implies $u = P
  u^{\prime}$ for some $u^{\prime} \in \mathcal{E}^{\prime} (
  \mathscr{D} )$.  Since, as a functional, $\Pi_{1} ( u )$ are defined
  on a set larger than $S$ we have that $\ker ( \Pi_{1} ) \subset
  \mathscr{K}$.  In order to obtain the opposite inclusion, let us
  define by $R$ the operator that realises the restriction on
  $\mathscr{C}_{p}^{+}$ and the subsequent multiplication by
  $\sqrt[4]{\abs{g_{AB}}}$.  Notice that $\Pi_{1}$ is defined as the
  composition $R \circ \Delta $, where now $\Delta$ is the map from
  $\mathcal{T}^{\prime} (\mathscr{D} )$ to $\mathcal{D}^{\prime} (
  \mathcal{O}^{\prime} )$, where $\mathcal{O}^{\prime}$ is the normal
  neighbourhood containing $\mathscr{D}$.  Hence $\ker ( \Pi_{1} )
  \supset \ker ( \Delta ) = \mathscr{K}$.  Note that $\mathscr{K}$ is
  contained in the ideal $\mathscr{I}$ divided out of $\mathscr{F}_{e}
  ( \mathscr{D} )$ in order to obtain $\mathscr{F}_{eo} (\mathscr{D}
  )$.  The proof can then be concluded by applying a similar procedure
  to $\Pi_{n}$ to verify that also $\ker ( \Pi_{n} )$ is contained in
  the ideal $\mathscr{I}$. \qed
\end{proof}

Before continuing the analysis of the map $\Pi$ acting on the extended
algebra $\mathscr{F}_{e} ( \mathscr{D} )$, we show that the pull-backs
both of the symplectic form $\sigma_{\mathscr{C}}$ and of the boundary
state $\omega$ have a nice interplay with the symplectic form in the
bulk and with the Hadamard states in general.  The next proposition
deals with the singular structure of the state $\omega$ when pulled
back in the bulk.
\begin{proposition}
  \label{Pro:Hadamard}
  Under the assumptions of Theorem~\ref{The:Pi1}
  \begin{equation}
    \label{eq:pull-back}
    H_{\omega} \doteq \Pi^{\ast} \omega
  \end{equation}
  is an Hadamard bi-distribution constructed as the pull-back of
  $\omega$ as in \eqref{eq:state} under $\Pi$ as in
  \eqref{eq:restriction}.
\end{proposition}
\begin{proof}
  The proof of this proposition can be performed by restricting
  attention to the compactly supported smooth functions on
  $\mathscr{D}$.  Let us start by showing that $H_{\omega}$ is a good
  distribution on $C^{\infty}_{0} ( \mathscr{D}^{2} )$, hence
  continuous in the topology of compactly supported smooth functions.
  To this end, notice that $H_{\omega} ( f , g ) = \omega ( \Pi_{1} (
  f ) , \Pi_{1} ( g ) )$, moreover, $\omega ( \Pi_{1} ( f ) , \Pi_{1}
  ( g ) )$ enjoys the continuity stated in \eqref{eq:2pcontinuity}.
  Furthermore, the $L^{2}$ norms present in \eqref{eq:2pcontinuity}
  are controlled by the supremum norms of $\Pi_{1} ( f )$, $\Pi_{1} (
  g )$ and their $r$-derivatives on some compact set in $\mathscr{C}$
  (here taken as the entire cone).  The proof of the continuity of
  $H_{\omega}$ can be concluded employing the continuity of $\Delta :
  C^{\infty}_{0} ( M ) \to C^{\infty} ( M )$, which is spoilt neither
  by the restriction on $\mathscr{C}^{+}_{p}$ nor by the
  multiplication by $\sqrt[4]{\abs{g_{AB}}}$.

  Notice that the antisymmetric part of $H_{\omega}$ equals the
  symplectic form \eqref{eq:symplbulk} which is preserved by the
  action of $\Pi$
  (Proposition~\ref{Pro:thesymplecticformisconserved}), and
  $H_{\omega}$ satisfies the equation weakly because so does $\Delta$
  which is used in the definition of $\Pi$.  Furthermore, $H_{\omega}
  ( f , f )$ is positive for every $f$, because $\omega$ is a state
  for the boundary algebra.  We thus conclude that $H_{\omega}$ is the
  two-point function of a quasi-free state for $\mathscr{F}_{b} (
  \mathscr{D} )$.  Hence, in order to prove the Hadamard property, due
  the the work of Radzikowksi \cite{Radzikowski:1996aa}, it is only
  necessary to check that the wave front set of $H_{\omega}$ satisfies
  the microlocal spectrum condition.  This can be verified following
  the procedure envisaged in \cite{Moretti:2008aa,hollands:2000}.  For
  completeness, we shall summarise here the main steps of such a
  proof.

\vskip .2cm

\noindent{\em 1.)} It suffices to show that the microlocal spectrum condition
    holds locally in $\mathscr{D}$, namely, when $H_{\omega}$ is
    restricted on a generic compact set $K^{2}$ with $K \subset
    \mathscr{D}$.  We hence have to show that
    \begin{equation}
      \label{eq:inclusionWF}
      \WF ( H_{\omega} ) = \big\{ ( x_{1} , \zeta_{1} ; x_{2} ,
      -\zeta_{2} ) \in T^{\ast} K^{2} \setminus \{ 0 \} \big\vert (
      x_{1} , \zeta_{1} ) \sim ( x_{2} , \zeta_{2} ) \text{,~}
      \zeta_{1} \triangleright 0 \big\} \text{,}
    \end{equation}
    where $\sim$ is the equivalence relation of \eqref{eq:WFDe}, while
    $\zeta_{1} \triangleright 0$ indicates that $\zeta_{1}$ is a
    future directed vector.  According to the preceding discussion, to
    show the inclusion $\supset$, we make use of Theorem~5.8 of
    \cite{Sahlmann:2001aa} which can be applied once $\subset$ holds
    and yields the other relation as thesis.
    
\vskip .1cm
    
\noindent{\em 2.)} In order to show that $\subset$ holds in
    \eqref{eq:inclusionWF}, notice that the past directed null
    geodesics originating from $K$ in $\mathcal{O}_{p}$ intersect
    $\mathscr{C}_{p}^{+}$ on a region contained in a compact set $N
    \subset \mathscr{C}_{p}^{+}$.  We stress that $p \not\in
    \mathscr{C}_{p}^{+}$, and hence $p \not\in N$.  Thus, if we
    smoothly localise $\hat{\Pi}_{1}$ (the integral kernel of
    $\Pi_{1}$) on $N^{\prime} \times K$, where $N^{\prime}$ is the
    complement of $N$ in $\mathscr{C}^{+}_{p}$, the resulting object
    is described by a smooth function which is square-integrable on
    $\mathscr{C}_{p}^{+}$ and so is its $V$-derivative, also when an
    entry of $\hat{\Pi}_{1}$ is kept fixed in $K$.
    
\vskip .1cm
    
\noindent{\em 3.)} We shall hence introduce a partition of unity on
    $\mathscr{C}_{p}^{+}$, $\Theta_{N} + \Theta^{\prime}_{N} = 1$,
    such that $\Theta_{N} \in C^{\infty}_{0} ( \mathscr{C}_{p}^{+} )$
    is equal to $1$ on the compact set $N$. Hence it vanishes on the
    intersection of a sufficiently small neighbourhood of $p$ with
    $\mathscr{C}_{p}^{+}$.  Inserting two such partitions of unity in
    $\Pi^{\ast} \omega$ and employing multilinearity, $H_{\omega}$
    becomes the sum of four terms, $\omega \big( ( \Theta_{N} +
    \Theta^{\prime}_{N} ) \hat{\Pi}_{1} \otimes ( \Theta_{N} +
    \Theta^{\prime}_{N} ) \hat{\Pi}_{1} \big)$.
    
\vskip .1cm
    
\noindent{\em 4.)}The only one which contributes to $\WF ( H_{\omega} )$ is
    $\omega ( \Theta_{N} \hat{\Pi}_{1} \otimes \Theta_{N}
    \hat{\Pi}_{1} )$.  In this case, we notice that, due to the form
    of $\WF ( \omega )$ given in \eqref{eq:WFomega} and to the
    constraint enjoyed by $\WF ( \hat{\Pi}_{1} )$ as discussed in 1)
    of Theorem~\ref{The:Pi1}, we can apply Theorem~8.2.13 of
    \cite{Hormander_The-Analysis-I:1990} in order to obtain that the
    inclusion sought holds for the wave front set of this term.
    
\vskip .1cm
    
\noindent{\em 5.)} All the other three terms have vanishing wave front sets.  Let
    us briefly discuss them separately.  In particular, due to the
    regularity shown by $\Theta^{\prime}_{N} \hat{\Pi}_{1}$ when
    restricted to $K$, the composition of $( \Theta^{\prime}_{N}
    \hat{\Pi}_{1} \otimes \Theta^{\prime}_{N} \hat{\Pi}_{1} )$ with
    $\omega$ on ${\mathscr{C}_{p}^{+}}^{2}$ can be computed and yields
    a smooth function on $K^{2}$.  The remaining two terms can be
    addressed similarly.  To this end, let us concentrate on $\omega (
    \Theta^{\prime}_{N} \hat{\Pi}_{1} \otimes \Theta_{N} \hat{\Pi}_{1}
    )$. At this point, notice that the supports of
    $\Theta^{\prime}_{N}$ and of $\Theta_{N}$ have non-vanishing
    intersection; however, such an intersection is contained in a
    further compact set $R$.  We can thus insert another partition of
    unity, $\Theta_{R} + \Theta^{\prime}_{R}$, in order to divide such
    a term in two parts $\omega (\Theta^{\prime}_{N} ( \Theta_{R} +
    \Theta^{\prime}_{R} ) \hat{\Pi}_{1} \otimes \Theta_{N}
    \hat{\Pi}_{1} \big)$.  Hence, $\omega \big( ( \Theta^{\prime}_{N}
    \Theta^{\prime}_{R} ) \hat{\Pi}_{1} \otimes \Theta_{N}
    \hat{\Pi}_{1} )$ has a vanishing wave front set because the
    supports of $\Theta^{\prime}_{N} \Theta^{\prime}_{R}$ and
    $\Theta_{N}$ are disjoint and $\omega$ is represented by a smooth
    function on their Cartesian product.  Furthermore, since both
    $\Theta^{\prime}_{N} \Theta_{R}$ and $\Theta_{N}$ are in
    $C^{\infty}_{0} ( \mathscr{C}^{+}_{p} )$, we can estimate the wave
    front set of $\omega \big( \Theta^{\prime}_{N} \Theta_{R}
    \hat{\Pi}_{1} \otimes \Theta_{N} \hat{\Pi}_{1} \big)$ employing
    once more Theorem~8.2.13 of \cite{Hormander_The-Analysis-I:1990}
    to obtain that it is equal to the empty set.
\end{proof}

We can now prove a second theorem which focuses on the effect of the
map $\Pi$ on the algebraic structures and on the boundary state.  In
particular, we shall individuate an Hadamard state in the bulk.
\begin{theorem}
  \label{The:Pi2}
  Under the assumptions of Theorem~\ref{The:Pi1}, one has:
  \renewcommand{\labelenumi}{\arabic{enumi})}
  \begin{enumerate}
  \item $\Pi$ induces a unit-preserving $^{\ast}$-homomorphism between
    the algebras $\big( \mathscr{F}_{e} ( \mathscr{D} ) ,
    \star_{H_{\omega}} \big)$ and $\big( \mathscr{A}_{e} (
    \mathscr{C}_{p}^{+} ) , \star_{\omega} \big)$.
  \item $\Pi$ is an injective $^{\ast}$-homomorphism when acting ``on
    shell'' on $\big( \mathscr{F}_{eo} ( \mathscr{D} ) ,
    \star_{H_{\omega}} \big)$.
  \end{enumerate}
\end{theorem}
\begin{proof}
  We only prove the first statement.  The second arises by direct
  inspection from this one and from Proposition~\ref{Pro:injectivity}.
  First notice that $\Pi$ automatically preserves the
  $^{\ast}$-operation because $\overline{\Pi}_{n} = \Pi_{n}$, hence
  \begin{equation*}
    \Pi_{n} ( F_{n} )^{\ast} = \Pi_{n} ( F_{n}^{\ast} ) \text{.}
  \end{equation*}
  Thus we only need to verify the statement on the $\star$-products.
  In particular, we look for $\star_{H_{\omega}} : \mathscr{F}_{e} (
  \mathscr{D} ) \times \mathscr{F}_{e} ( \mathscr{D} ) \to
  \mathscr{F}_{e} ( \mathscr{D} )$ such that,
  \begin{equation}
    \label{eq:compatibleproduct}
    \Pi ( F \star_{H_{\omega}} G ) = ( \Pi F ) \star_{\omega} ( \Pi G
    ) \text{,} \quad \forall F \text{,~} G \in \mathscr{F}_{e} (
    \mathscr{D} ) \text{,}
  \end{equation}
  and, at the same time $\big( \mathscr{F}_{e} , \star_{H_{\omega}}
  \big)$ is isomorphic to $\big( \mathscr{F}_{e} , \star_{H} \big)$.

  The natural candidate arises from the analysis performed in
  Proposition~\ref{Pro:Hadamard} and, in particular, from the
  distribution $H_{\omega}$ introduced in \eqref{eq:pull-back} to be
  plugged in \eqref{eq:starproduct} in place of $H$.  This is a
  well-defined procedure since $H_{\omega}$ is of Hadamard form as
  proved in Proposition~\ref{Pro:Hadamard}, and, hence, $(
  \mathscr{F}_{e} , \star_{H} )$ turns out to be isomorphic to $(
  \mathscr{F}_{e} , \star_{H_{\omega}} )$, the isomorphism being
  realised as in \eqref{eq:isomorphism}.  We are thus left with the
  task to verify \eqref{eq:compatibleproduct} for every $F$ and $G$ in
  $\mathscr{F}_{e} ( \mathscr{D} )$.  If we exploit both the
  definitions and the bilinearity of all the $\star$-products
  involved, this reduces to the requirement to show that
  \begin{equation*}
    \Pi_{l+m-2k} \big( ( F_{l} \otimes G_{m} ) ( {H_{\omega}}^{\otimes
      k} ) \big) = ( \Pi_{l} F_{l} \otimes \Pi_{m} G_{m} ) (
    \omega^{\otimes k} )
  \end{equation*}
  for $l + m - 2 k \geqslant 0$.  The last relation directly results
  from bilinearity, \eqref{eq:pull-back} and from the fact that
  $\Pi_{k} = \Pi_{k_1} \otimes \Pi_{k_2}$ for all $k_{1}$, $k_{2} > 0$
  such that $k_{1} + k_{2} = k$.  \qed
\end{proof}

\begin{remark}
  Notice that it is possible to turn the injective homomorphism into a
  bijection if we restrict attention to the local von Neumann algebras
  defined as the double commutant in the GNS representation of a
  quasifree Hadamard state of the $C^{\ast}$-algebra generated by the
  local Weyl operators constructed out of the symplectic forms
  \eqref{eq:symplbulk} and \eqref{eq:symplecticform} in $\mathscr{D}$
  and on the boundary $\mathscr{C}_{p}^{+}$, respectively.  This last
  claim is based on the invertibility of $\Pi_{2}$ on the weak
  solutions of the Klein-Gordon equation, 
where we recall
  that the Goursat problem with compact initial data on
  $\mathscr{C}_{p}$, in general, yields a solution of \eqref{eq:eqm}
  whose restriction on any Cauchy surface of $\mathscr{D}$ is not
  compact.

  Alas, the von Neumann algebra mentioned does not contain relevant
  physical observables such as the components of the regularised
  stress-energy tensor or the regularised squared fields, objects we
  would like to use in order to extract information about the local
  geometric data such as the scalar curvature.
\end{remark}

\section{Interplay with General Covariance and Comparison between
  Spacetimes}
\label{sec:4}

We are now in the position to collect all our results in a
comprehensive framework which will exhibit both a nice interplay with
the principle of local covariance, as devised in
\cite{Brunetti:2003aa}, and the possibility to compare quantum field
theories on different backgrounds both at the level of algebras and of
states.

To this avail, the construction in Subsection~\ref{ssec:3.5} will play
a pivotal role, and the natural language we adopt is that of
categories, which was already introduced in Subsection~\ref{ssec:2.1}.
In particular, notice that the construction of an extended algebra of
observables on a double cone $\mathscr{D}$ can be realised as a
suitable functor between the categories $\mathsf{DoCo}_{iso}$ and
$\mathsf{Alg}$, although it is not possible to extend such a functor
to $\mathsf{DoCo}$.

Notwithstanding this obstruction, the additional structure which
arises from both the boundary and the field theory defined thereon
allows us to circumvent the above problem in a way that also puts us
in a position to compare field theories in different spacetimes.  This
requires the introduction of a further category, namely,
\begin{itemize}
\item[$\mathsf{BAlg}$:] The objects of this category are the extended
  boundary (topological $^{\ast}$-)algebras presented in
  Subsection~\ref{ssec:3.4}, constructed on all possible
  $\mathscr{C}_{p}^{+}$, while the morphisms are the unit preserving
  $^{\ast}$-homomorphisms among them.
\end{itemize}

The key point consists in making a profitable use of the
$^{\ast}$-homomorphisms $\Pi$ introduced in Theorem \ref{The:Pi1}, in
order to establish that $\Pi \circ \mathscr{F}$ indeed defines a
functor between the two categories $\mathsf{DoCo}$ and
$\mathsf{BAlg}$, which admits the following pictorial, but inspiring
diagrammatic representation,
\begin{equation}
  \label{eq:commutativediagram}
  \begin{CD}
    \mathscr{D} @>{\mathscr{F}}>>  \mathscr{F}_{e} ( \mathscr{D} )
    @>{\Pi}>>\mathscr{A}_{e} ( \mathscr{C}^{+}_{p} ) \\
    @V{\imath_{e,e^{\prime}}}VV @.
    @VV{\alpha_{\imath_{e,e^{\prime}}}}V\\ 
    \mathscr{D}^{\prime} @>>{\mathscr{F}}> \mathscr{F}_{e} (
    \mathscr{D}^{\prime} ) @>>{\Pi^{\prime}}>\mathscr{A}_{e} (
    {\mathscr{C}^{+}_{p}}^{\prime} )
  \end{CD}
  \qquad \raisebox{-2\baselineskip}{\text{.}}
\end{equation}
The arrow $\alpha_{\imath_{e,e^{\prime}}}$ traces back to the analysis
in Subsection~\ref{ssec:3.2}, where it was shown that the boundary
theory can be constructed and analysed independently of the specific
bulk.  Hence, $\alpha_{\imath_{e,e^{\prime}}}$ is the
$^{\ast}$-homomorphism $\alpha_{\imath_{e,e^{\prime}}} :
\mathscr{A}_{e} ( \mathscr{C}^{+}_{p} ) \to \mathscr{A}_{e} (
{\mathscr{C}^{+}_{p}}^{\prime} )$ whose action on the $F^{\prime} \in
\mathscr{A}_{e} ( \mathscr{C}^{+}_{p} ) $ is defined as follows: By
means of the push-forward, it is
\begin{equation*}
  \alpha_{\imath_{e,e^{\prime}}} ( F^{\prime}_{n} ) =
  {\imath_{e,e^{\prime}}}_{\ast} F^{\prime}_{n} \text{,}
\end{equation*}
on $\big( \imath_{e,e^{\prime}} ( \mathscr{C}^{+}_{p} ) \big)^{n}
\subset ( {{\mathscr{C}^{+}_{p}}^{\prime}} )^{n}$, while
$\alpha_{\imath_{e,e^{\prime}}} ( F^{\prime}_{n} ) = 0$ on the
complement of $\big( \imath_{e,e^{\prime}} ( \mathscr{C}^{+}_{p} )
\big)^{n}$ in $( {\mathscr{C}^{+}_{p}}^{\prime} )^{n}$.  Such an
operation is well-defined because $F^{\prime}_{n}$ has compact support
towards the future and, when extended on the closure of
$\mathscr{C}^{+}_{p}$, $\imath_{e,e^{\prime}}$ maps $p$ to
$p^{\prime}$.  We hence have the following
\begin{proposition}
  Consider $\mathscr{A}_{e} : \mathsf{DoCo} \to \mathsf{BAlg}$, whose
  action on the objects and morphisms is seen as follows,
  \begin{itemize}
  \item the action of $\mathscr{A}_{e}$ on the objects of
    $\mathsf{DoCo}$ is such that $\mathscr{A}_{e} ( \mathscr{D} ) =
    \Pi \circ \mathscr{F}_{e} ( \mathscr{D} ) = \mathscr{A}_{e} (
    \mathscr{C}_{p}^{+} )$;
  \item the action of $\mathscr{A}_{e}$ on the morphisms
    $\imath_{e,e^{\prime}}$ is such that $\mathscr{A}_{e} (
    \imath_{e,e^{\prime}} )= \alpha_{\imath_{e,e^{\prime}}}$.
  \end{itemize}
  Then $\mathscr{A}_{e}$ is a functor between the two categories.
\end{proposition}
\begin{proof}
  In order to show that $\mathscr{A}_{e} : \mathsf{DoCo} \to
  \mathsf{BAlg}$ is a functor, notice that the covariance property
  holds and the identity is preserved, \textit{i.e.},
  \begin{equation*}
    \mathscr{A}_{e} ( \imath_{e,e^{\prime}} ) \circ \mathscr{A}_{e} (
    \tilde{\imath}_{e^\prime,\tilde{e}^{\prime}} ) = \mathscr{A}_{e}
    ( \imath_{e,e^{\prime}} \circ
    \tilde{\imath}_{e^\prime,\tilde{e}^{\prime}} ) \text{,} \qquad
    \mathscr{A}_{e} ( id_{\mathscr{D}} ) = id_{\mathscr{A}_{e} (
      \mathscr{C}_{p}^{+})} \text{,}
  \end{equation*}
  as can be seen by direct inspection of the definition of
  $\mathscr{A}_{e} ( \imath_{e,e^{\prime}} )$.
  \qed
\end{proof}

As for the connection with the principle of general local covariance,
we recall that, in the most general case, it is not possible to find a
direct relation between $\mathscr{F} ( \mathscr{D} )$ and $\mathscr{F}
( \mathscr{D}^{\prime} )$, unless the embedding $\Pi : \mathscr{F} (
\mathscr{D}^{\prime} ) \to \mathscr{F} ( \mathscr{C}^{\prime} )$ can
be inverted on the image of $\Pi$ composed with
$\alpha_{\imath_{e,e^{\prime}}}$.  This is indeed what happens
whenever, \emph{e.g}, $\imath_{e,e^{\prime}}$ is an isometry (or, at
worst, a conformal isometry \cite{Pinamonti:2009aa}) which preserves
the base point $p$.  Hence we are working in $\mathsf{DoCo}_{iso}$.

Under this assumption, the discussion about causality, usually an
integral part of the reasoning as in
\cite{Brunetti:2003aa,Brunetti:2009aa}, does not have to be performed
directly, since its essence is already encoded in the analysis of the
properties of the map $\Pi$.  A similar statement holds true also for
the time-slice axiom (see \cite{Chilian:2009aa}, in particular).
Especially, since the theory on the boundary is, to a certain extent,
non-dynamical, there is no such axiom in our boundary framework and
the one in the bulk is automatically assured by $\Pi$ and its
properties.

\subsection{Comparison of Expectation Values in Different Spacetimes}
\label{ssec:4.1}

The aim of this section is to clarify in which sense one can compare
two field theories on two different backgrounds.  We shall first
explain the procedure abstractly and then give a concrete example.

Bearing in mind \eqref{eq:commutativediagram}, it is straightforward
to realise that, whenever we assign a state $\omega$ on
$\mathscr{A}_{e} ( \mathscr{C}^{\prime}_{p} )$, we can pull it back
either on $\mathscr{F}_{e} ( \mathscr{D}^{\prime} )$ via
$\Pi^{\prime}$ or on $\mathscr{F}_{e} ( \mathscr{D} )$ via
$\alpha_{\imath_{e,e^{\prime}}} \circ \Pi$, and the information about
the bulk geometry is indeed restored by $\Pi$ and $\Pi^{\prime}$.
This is a rather general feature which holds true regardless of the
global structure of the spacetimes in which $\mathscr{D}$ and
$\mathscr{D}^{\prime}$ are embedded.  Yet, on practical grounds, it is
natural to choose one of the two double cones embedded in the
four-dimensional Minkowski spacetime, where our capability of
performing explicit computations of physical quantities is enhanced
due to the large symmetry of the background.

To be more specific on this point, consider a double cone
$\mathscr{D}$ as a subset of $( \mathbb{R}^{4} , \eta )$ while
$\mathscr{D}^{\prime}$ lies in a generic strongly causal spacetime
$M^{\prime}$, chosen in such a way that there exist two frames $e$ and
$e^{\prime}$ in $T_{p} ( \mathbb{R}^{4} )$ and $T_{p^{\prime}} (
M^{\prime} )$, respectively, yielding a well-defined
$\imath_{e,e^{\prime}} : \mathscr{D} \to \mathscr{D}^{\prime}$.  At
this point we can apply the construction discussed for the local
fields and algebras, related on the boundary via the map
$\alpha_{\imath_{e,e^{\prime}}}$.

As a next step, following also the general philosophy of
\cite{Brunetti:2003aa}, we consider observables constructed out of the
same local fields either on $\mathscr{D}$ or $\mathscr{D}^{\prime}$.
Here we suppose that there exists $\imath_{e,e^{\prime}} : \mathscr{D}
\to \mathscr{D}^{\prime}$ and hence, from the same field, we form
$\Phi ( f ) \in \mathscr{F}_{e} ( \mathscr{D} )$ and $\Phi (
{\imath_{e,e^{\prime}}}_{\ast} f ) \in \mathscr{F}_{e} (
\mathscr{D}^{\prime} )$, where $\Phi$ is a local field in the sense of
\cite{Brunetti:2003aa}.  We compute their expectation values on the
pull-back of a suitable boundary state yielding an Hadamard
counterpart in the bulk.  In general, the difference between the
results depends on the geometric data of both $\mathscr{D}$ and
$\mathscr{D}^{\prime}$, and thus we are ultimately comparing quantum
field theories on different backgrounds.

Nonetheless, from a computational point of view, this procedure is
still too involved, since one has to cope with the singular structure
of the chosen state(s).  Even if we restrict attention to those
fulfilling the Hadamard condition, one would still need to take care
of the regularisation procedure of the observables, a hassle which can
be avoided.  The idea is to consider on each double cone two bulk
Hadamard states constructed out of the pull-back of different boundary
counterparts and to work with their difference.  In this case the
integral kernel of such a difference is known to be smooth, an
advantage which strongly reduces the computational efforts.  Although
the price to pay is the introduction of two states on the light cone,
a natural candidate as one of them is the distinguished reference
state $\omega$ which arises from \eqref{eq:pull-back} in
Proposition~\ref{Pro:Hadamard}.  Hence we are left with the need to
assign only one extra datum.

Before we discuss an explicit example of this procedure, we stress the
most important properties of both $\omega$ and of its pull-back in the
bulk, say $\Pi^{\ast} \omega$.  These can be inferred from both
Proposition~\ref{Pro:stateomega} and Theorem~\ref{The:Pi2}:
\renewcommand{\labelenumi}{\arabic{enumi}.}
\begin{enumerate}
\item \textit{Local Lorentz invariance}: According to the third item
  of Proposition~\ref{Pro:stateomega}, $\omega$ turns out to be
  invariant under a set of geometric transformations on the full
  cone $\mathscr{C}$ which contains the boundary.  
  In particular,
  $\omega$ is always invariant under the natural action of the subgroup of the Lorentz group which
corresponds to isometries of the neighbourhood where the state is defined.
 Since the map $\Pi$ is constructed substantially
  out of the causal propagator \eqref{eq:causalp} in a geodesic
  neighbourhood, its action on $\omega$ via pull-back does not spoil
  the above property, \textit{i.e.}, $\Pi^{\ast} \omega$ is invariant
  under the above assumptions.
\item \textit{Microlocal structure}: The wave front set of $\omega$ is
  contained in the union of \eqref{eq:WFA} and \eqref{eq:WFB} and,
  most notably, it does not contain directions to the past.  In
  particular, this allows for the proof of
  Proposition~\ref{Pro:Hadamard} according to which $\Pi^{\ast}
  \omega$ satisfies the Hadamard property in the bulk double cone.
\item \textit{Behaviour as a ``vacuum''}: The boundary state $\omega$
  turns out to be invariant under rigid translations of the
  $V$-coordinate, or, in other words, it is a vacuum with respect to
  the transformation generated by the vector $\partial_{V}$.  This
  statement can be proved exactly as in
  \cite{Moretti:2006aa,Dappiaggi:2009ab} for the counterpart on the
  conformal boundary of an asymptotically flat or of a cosmological
  spacetime, \textit{viz.}, from the explicit form of the two-point
  function \eqref{eq:L2integral}.  This also entails that the energy
  computed on the cone with respect to $\partial_{V}$ is minimised.
  Unfortunately, this property has not a strong counterpart in the
  bulk, but, if the bulk can be realised as an open set in $(\mathbb{R}^{4} , 
\eta )$, then $\Pi^{\ast} \omega$ is seen to
coincide with the Minkowski vacuum for massless fields.
\end{enumerate}

\subsection{An Application: Extracting the Curvature}
\label{ssec:4.2}

In this subsection we shall present an explicit application which
follows the guidelines given above.  For simplicity consider on the
one hand a double cone $\mathscr{D}$ realised as an open subset of
Minkowski spacetime $( \mathbb{R}^{4} , \eta )$, where the metric
$\eta$ has the standard diagonal form with respect to the Cartesian
coordinates $( t , x , y , z )$ induced by the standard orthonormal
frame $e$ of $\mathbb{R}^{4}$.  As $\mathscr{D}^{\prime}$ we consider
a double cone which can be embedded in a homogeneous and isotropic
solution of Einstein's equation with flat spatial section.  This is a
Friedman-Robertson-Walker spacetime $( M^{\prime} , g )$, where $g =
a^{2} ( t ) \thinspace \eta$ and $a ( t ) \in C^{\infty} ( I ,
\mathbb{R}^{+} )$ with $I \subseteq \mathbb{R}$ and $a ( 0 ) = 1$.
Here $t$ refers to the so-called conformal time and thus we still
consider the coordinates $( t , x , y , z )$ induced by the standard
frame of $\mathbb{R}^{4}$, indicated as $e^{\prime}$ to distinguish it
from the previous one.  Furthermore, notice that, in view of the
special form of $g$, $e^{\prime} = \frac{e}{a ( t )}$.

Since the underlying spacetimes are conformally related, their causal
structures and, in particular, the double cones coincide.  Consider
two points $p = ( 0 , 0 , 0 , 0 )$ and $q = ( t^{\prime} , 0 , 0 , 0
)$ and the corresponding double cones $\mathscr{D} ( p , q ) \subset
\mathbb{R}^{4}$ and $\mathscr{D}^{\prime} ( p , q ) \subset
M^{\prime}$.  In this framework the map $\imath_{e,e^{\prime}} :
\mathscr{D} ( x_{0} , x_{1} ) \to \mathscr{D}^{\prime} ( x_{0} , x_{1}
)$ turns out to be trivial.  Next, we choose a minimally coupled real
scalar field theory, \textit{viz.},
\begin{equation*}
  \phi : \mathscr{D} \to \mathbb{R} \text{,} \quad \Box \phi = 0
  \text{,}
\end{equation*}
where $\Box$ is the d'Alembert wave operator constructed out of the
metric in $\mathscr{D}$.  We stress once more that we consider the
very same equation also in $\mathscr{D}^{\prime}$.  If we follow the
guidelines of Subsection~\ref{ssec:3.1}, we can construct 
$\mathscr{F}_{e}(\mathscr{D})$ and $\mathscr{F}_{e}(\mathscr{D}^{\prime})$ 
and their counterparts on the boundaries
$\mathscr{C}_{p}$ and $\mathscr{C}^{\prime}_{p}$.  As outlined in the
previous subsection, we now consider two algebraic states on
$\mathscr{A}_{e} ( \mathscr{C}_{p} )$, one, $\omega$, is the reference
state, while the other can be arbitrary, provided that the pull-back
to the bulk via $\Pi$ still fulfils the Hadamard condition.  This
requirement is not too restrictive since, \textit{e.g.}, any state
which differs from $\omega$ by a smooth function on the boundary and
vanishes in a neighbourhood of the tip is an admissible choice.

In particular, consider another Gaussian state $\omega^{\prime} :
\mathscr{A}_{e} ( \mathscr{C}_{p} ) \to \mathbb{C}$, whose two-point
function has the following form,
\begin{equation}
  \label{eq:secondstate}
  \omega^{\prime} ( \psi_{1} , \psi_{2} ) = \omega ( \psi_{1} ,
  \psi_{2} ) + \frac{1}{4\pi} \int\limits_{\mathbb{R}^{+} \times
    \mathbb{R}^{+} \times \mathbb{S}^{2}} dr \thinspace dr^{\prime}
  d\mathbb{S}^{2} ( \theta , \varphi ) \medspace \psi_{1} ( r , \theta
  , \varphi ) \psi_{2} ( r^{\prime} , \theta , \varphi ) \text{.}
\end{equation}
Notice that the integral on the right-hand side entails that the
integral kernel of $\omega^{\prime} - \omega$ is not smooth because it
contains a $\delta$-like singularity in the angular coordinates.
Despite this fact $\omega^{\prime}$ can be pulled back to every
spacetime still yielding an Hadamard bi-distribution.

The last statement can be proved operating in the same way as in the
proof of Proposition~\ref{Pro:Hadamard} with $\omega$ replaced by
$\omega^{\prime} - \omega$.  The key new feature, yielding $\WF \big(
H_{\omega^{\prime} - \omega} \big) = \emptyset$, is the fact that,
while $( \zeta_{x} )_{r}$ is never equal to zero for every $( x ,
\zeta_{x} ; y , \zeta_{y} ) \in \WF ( \Pi_{1} )$, every $( x ,
\zeta_{x} ; y , \zeta_{y} ) \in \WF ( \omega^{\prime} - \omega )$ is
such that $( \zeta_{x} )_{r} = ( \zeta_{y} )_{r} = 0$.

We are now ready to consider the expectation values of suitable
observables from the point of view of both $\mathscr{D}$ and
$\mathscr{D}^{\prime}$.  The most natural one is the expectation value
of $\phi^{2} ( f )$, where $f \in C^{\infty}_{0} ( \mathscr{D} )$ (and
hence also in $C^{\infty}_{0} ( \mathscr{D}^{\prime} )$), with respect
to the bulk states constructed out of $\omega^{\prime}$ and $\omega$.
Notice that $\phi^{2} ( f )$ is a shortcut for saying that we are
actually considering $u_{f} = f ( x ) \delta ( x - y ) \in
\mathcal{A}_{s}^{2} ( \mathscr{D} ) \subset \mathscr{F}_{e} (
\mathscr{D} )$.  One of the advantages of this construction is that,
in this case, we are allowed to keep a more general stance, namely, we
can substitute $f ( x )$ by a Dirac function peaked at the point
$x_{t} = ( t , 0 , 0 , 0 )$ with $t \in ( 0 , t^{\prime} )$.  This is
tantamount to consider $: \phi^{2} : ( x_{t} )$, where $u = \delta (
x - x_{t} ) \delta ( x - y ) \in \mathscr{A}_{s}^{2} ( \mathscr{D}
)$.

Now \eqref{eq:restriction} can be used to evaluate $\Pi_{2} u$ in both
$( \mathbb{R}^{4} , \eta )$ and $( M^{\prime} , g )$ by means of the
explicit form of the causal propagator.  In Minkowski spacetime it
looks like (see \cite{Friedlander_The-Wave-Equation:1975} or
\cite{Poisson:2004ab})
\begin{equation}
  \label{Minkp}
  \Delta ( x , x^{\prime} ) \doteq - \frac{\delta ( t - t^{\prime}
    - \abs{{\bf x} - {\bf x^{\prime}}} )}{4 \pi \abs{{\bf x}
      - {\bf x^{\prime}}}} + \frac{\delta ( t - t^{\prime} +
    \abs{{\bf x} - {\bf x^{\prime}}} )}{4 \pi \abs{{\bf x} -
      {\bf x^{\prime}}}} \text{,}
\end{equation}
where $t$ is the time coordinate and ${\bf x}$ the
three-dimensional spatial vector in Euclidean coordinates.  The
counterpart of $\Delta$ in $\mathscr{D}^{\prime}$ can be directly
evaluated exploiting the conformal transformation between $(
M^{\prime} , g )$ and $( \mathbb{R}^{4} , \eta )$.  The d'Alembert
wave equation in the first spacetime $( M^{\prime} , g )$ corresponds
in the flat one to
\begin{equation}
  \label{eq:confequiv}
  \Box \phi - \frac{a^{\prime\prime}}{a} \phi = 0 \text{,}
\end{equation}
where $^{\prime}$ stands for time derivation and $\Box = -
\nabla^{\mu} \nabla_{\mu}$.  Furthermore, the causal propagator
  $\widetilde{\Delta}$ of this partial differential equation is
  related to the one in $M^{\prime}$ via
\begin{equation}
  \label{rel1}
  \Delta_{M^{\prime}} ( x , y ) = \frac{1}{a ( t_{x} ) a ( t_{y} )}
  \widetilde{\Delta} ( x , y ) \text{.}
\end{equation}
If we follow the procedure discussed in Chapter~4 of
\cite{Poisson:2004ab}, we get
\begin{equation*}
  \widetilde{\Delta} ( x , x^{\prime} ) = \Delta ( x , x^{\prime} ) +
  \frac{V ( x , x^{\prime} )}{4 \pi} \big( \Theta ( t - t^{\prime} -
  \abs{{\bf x} - {\bf x^{\prime}}} ) - \Theta ( -t + t^{\prime} -
  \abs{{\bf x} - {\bf x^{\prime}}} ) \big) \text{,}
\end{equation*}
where $V ( x , x^{\prime} )$ is a smooth function whose explicit form
is derived from the Hadamard recursive relations for
\eqref{eq:confequiv}.  In particular, it also holds that $V ( x , x )
= -\frac{a^{\prime\prime} ( x )}{2a ( x )}$.

We are now ready to compare the expectation values in $(
\omega^{\prime} - \omega )$.  The Minkowski side of this operation
yields, by direct computation,
\begin{equation*}
  ( \omega^{\prime} - \omega ) ( \Pi_{2}^{\mathbb{R}^{4}} u ) =
  \frac{1}{4} \text{,}
\end{equation*}
while in the cosmological setting
$$
(\omega'-\omega)(\Pi_2^{M'}u)= 
\left[
(4\pi)\int_0^{\infty} \frac{\widetilde{\Delta}(x_t,r_*)}{a(t)a(r_*)} r_* a(r_*)   a(r_*)^2  dr_* 
\right]^2, 
$$
where we have rewritten the integral in the $r$-variable in terms of $r_*$, the affine parameter of the null cone in Minkowski spacetime. The defining relation between the two variables is
$$
dr = a^2(r_*) dr_*.
$$
The above integral can be rewritten by means of \eqref{rel1} as
$$
(\omega'-\omega)(\Pi_2^{M'}u)= 
\left[
4\pi \int_0^{\infty}  {\Delta}(x_t,r_*) \frac{a(r_*)^2}{a(t)} r_* dr_*   
+
\int_0^{t/2} V(x_t,r_*) \frac{a(r_*)^2}{a(t)} r_* dr_* 
\right]^2,
$$
in which the first integral yields via \eqref{Minkp}
$$
(\omega'-\omega)(\Pi_2^{M'}u)= 
\left[
 \int_0^{\infty}  \delta(t-2 r_*)
\frac{a(r_*)^2}{a(t)} dr^*   
+
\int_0^{t/2} V(x_t,r_*) \frac{a(r_*)^2}{a(t)} r_* dr_* 
\right]^2.
$$
Let us now expand $a(t)$ in a power series around the point $x_0$,
$$
(\omega'-\omega)(\Pi_2^{M'}u)= 
\left[
\frac{1}{2} 
\frac{a(t/2)^2}{a(t)}   
+
a''(x_0) \frac{t^2}{4} + O(t^3)
\right]^2, 
$$
where, in the derivative, we have exploited that, at first order in $t$,
$$
V(x_t,r_*)= \frac{a''(x_0)}{a(x_0)} + O(t),
$$
also due to the rotational symmetry of $M'$. If we now expand both $a(t)$ and $a(t/2)$ in a Taylor series , we obtain
\begin{gather*}
(\omega'-\omega)(\Pi_2^{M'}u)= 
\left[
\frac{a}{2} 
\left[
1+\left( -\frac{1}{2}\frac{a''}{a}+ \frac{3}{2} \left(\frac{a'}{a}\right)^2  \right) t^2
\right]
+
a'' \frac{t^2}{4}+ O(t^3)
\right]^2=\\
\left[
\frac{a}{2}+\frac{3 a}{4}\left(\frac{a'}{a}\right)^2 t^2 +O(t^3)
\right]^2,
\end{gather*}
where all functions $a$ together with their derivatives are evaluated at $x_0$.
We can summarize the discussion, finally calculating the difference between $(\omega'-\omega)(\Pi_2^{\mathbb{R}^4}u)$ and $(\omega'-\omega)(\Pi_2^{M'}u)$,
$$
(\omega'-\omega)(\Pi_2^{\mathbb{R}^4}u)-(\omega'-\omega)(\Pi_2^{M'}u)= \frac{3}{4}(a'(x_0))^2 t^2 + O(t^3),
$$
with $a(x_0)=1$. The interpretation of this result is that, exactly as expected, the above comparison yields a result which, at first order, allows us to extract via a measurement precise information on the a priori unknown geometric data, in this case, the derivative of the scale factor at the point $x_0$ in a Friedman-Robertson-Walker universe. 

\section{Summary and Outlook}
\label{sec:5}

In this paper we have achieved a twofold goal: on the one hand we
propose a novel way to look at the properties of a local quantum field
theory in a suitable curved background, while, on the other hand, the
very same construction yields a mechanism which allows for the
comparison of expectation values of field observables in different
spacetimes.

More specifically, starting from a careful analysis of the underlying
geometry, we realise that only moderate assumptions are needed to
reach our goals, \textit{viz.}, our general setting consists of an
arbitrary strongly causal manifold $M$ in which we identify an
arbitrary but fixed double cone $\mathscr{D} \equiv \mathscr{D} ( p ,
q ) = I^{+} ( p ) \cap I^{-} ( q )$ strictly contained in a normal
neighbourhood of $p$.  Since $\mathscr{D}$ is globally hyperbolic, we
can consider therein a real scalar field theory along the lines of
\eqref{eq:eqm} and therefore follow the general quantisation scheme
which particularly calls for the association of a Borchers-Uhlmann
algebra of observables with the chosen system.  This algebra can be
extended, both enlarging the set of its elements and the defining
product, in order to encompass also \textit{a priori} more singular
objects, such as the Wick polynomials, which constitute the so-called
extended algebra.  The very deep reason for choosing $\mathscr{D}
\subset M$ lies in its boundary and, more properly, on the portion of
$J^{+} ( p )$ which it contains.  This is a differentiable submanifold
of codimension $1$ on which it is possible to construct a genuine free
scalar field theory, following exactly the same procedure successfully
employed for the causal boundary of an asymptotically flat or
cosmological spacetime in \cite{Dappiaggi:2006aa,Dappiaggi:2009aa}.
The main novel result in this framework arises from the construction
of an extended algebra also for the boundary theory---$\mathscr{A}_{e}
( \mathscr{C}_{p} )$ in the main body---whose well-posedness is
justified both by its mathematical properties and by its relation to
the bulk counterpart.  Hence, the latter is embedded in
$\mathscr{A}_{e} ( \mathscr{C}_{p} )$ by means of $\Pi$, an injective
$^{\ast}$-homomorphism.

The advantage of this picture is the possibility to make use of a long
tradition, originating from \cite{Kay:1991aa}, which allows us to
exploit the geometrical properties of the boundary to identify for the
algebra thereon a natural state which can be pulled-back to the bulk
via $\Pi$, yielding a counterpart which satisfies the microlocal
spectrum condition, hence is of Hadamard form.  This guarantees that
we can identify a local state in $\mathscr{D}$ which is physically
well-behaved.  In physical terms it means that this state is the same for all
inertial observers at $p$.  In other words, the bulk state as well as
the one on the boundary are invariant under a natural action of
$SO_{0} (3,1)$.  Thus it can be identified as a sort of local vacuum
on a curved spacetime independent of the frame.  

The second goal of comparing expectation values on different
backgrounds is based on the above construction.  More precisely, we
consider not just one but actually two regions as above in two
\textit{a priori} different spacetimes $M$ and $M^{\prime}$.  The
construction of the field theories proceeds as usual, but now we make
use of the invertibility of the exponential map in geodesic
neighbourhoods in order to engineer the double cones $\mathscr{D}
\subset M$ and $\mathscr{D}^{\prime} \subset M^{\prime}$ so that we
can map the boundary in $\mathscr{D}$ to that in
$\mathscr{D}^{\prime}$ via a local diffeomorphism.  This procedure can
be brought to the level of boundary extended algebras which thus can
be related by means of a suitable restriction homomorphism.  The
advantage is the previously unknown possibility to carefully use the
distinguished state identified on each $\mathscr{C}_{p}$ in order to
compare the bulk expectation values of field observables constructed
for the theories in $\mathscr{D}$ and $\mathscr{D}^{\prime}$.  The
important point is that this new perspective is completely compatible
with the standard principle of general local covariance when
applicable as devised in \cite{Brunetti:2003aa}, and, actually, it
complements it corroborating its significance.

Furthermore, since nothing prevents us from choosing one of the
spacetimes as the Minkowski one, one can concretely check how the
proposed machinery allows for the comparison of the expectation values
of the field observables, making manifest the role and the magnitude
of the geometric quantities.  We stress this point by means of a
simple example involving a massless minimally coupled field in the
flat and in a cosmological spacetime.  It seems safe to claim that
there are several possibilities to apply our procedure to many other
cases of physical interest.  These are certainly not the only roads
left open, and actually even the identified bulk Hadamard state should
be studied in more detail.  As a matter of fact, it is interesting to
understand whether it is connected in any way with the states of
minimum energy that appear in Friedman-Robertson-Walker spacetimes
\cite{Olbermann:2007aa}.  We leave this as well as the myriad of other
questions for future investigations.

\section*{Acknowledgements.} 
  The work of C.D. is supported by a Junior Research Fellowship from
  the Erwin Schr\"odin\-ger Institute and he gratefully acknowledges it,
  while that of N.P. is supported by the German DFG Research Program
  SFB 676.  We would like to thank Bruno Bertotti, Romeo Brunetti,
  Klaus Fredenhagen, Thomas-Paul Hack and Valter Moretti for
  profitable discussions on the topics of the present paper.
  C.D. also gratefully acknowledges the support of the National
  Institute for Theoretical Physics of South Africa for his stay at
  the University of KwaZulu-Natal.

\appendix
\section{Hadamard States}
\label{sec:a}

This appendix briefly recollects some properties of Hadamard states
which are used throughout the main text.  Since most of the material
has already been proved in several different alternative ways in the
literature, we limit ourselves to giving the main statements and the
necessary references.  Let us stress that, from a physical
perspective, Hadamard states are the natural candidates for physical
ground states of a quantum field theory on a curved background, since
their ultraviolet behaviour mimics that of the Minkowski vacuum at
short distances and, furthermore, they guarantee that the quantum
fluctuations of the expectation values of observables, such as the
smeared components of the stress-energy tensor, are finite.

In the subsequent discussion we always assume that we are dealing with
a quasi-free state on a suitable field algebra constructed on a
globally hyperbolic spacetime $( M , g )$ from a field satisfying an
equation of motion such as \eqref{eq:eqm}.  We stick to this
assumption because it is consistent with the main body of the paper,
but the reader should keep in mind that such an hypothesis could be
relaxed (see for example \cite{Sanders:2010aa}).  As a starting point
we state a global criterion characterising Hadamard states
\cite{Radzikowski:1996aa,Radzikowski:1996ab}.
\begin{definition}
  \label{Def:globHad}
  A state $\omega$ satisfies the \textbf{Hadamard condition} and is
  thus called an \textbf{Hadamard state} if and only if
  \begin{equation*}
    \WF ( \omega ) = \big\{ ( x , k_{x} ; y , -k_{y} ) \in T^{\ast}
    M^{2} \setminus \{ 0 \} \big\vert ( x , k_{x} ) \sim ( y , k_{y} )
    \text{,~} k_{x} \triangleright 0 \big\} \text{,}
  \end{equation*}
  where, in this expression, $\omega$ actually stands for the integral
  kernel of the two-point function associated with $\omega$.  The
  relation $( x , k_{x} ) \sim ( y , k_{y} )$ indicates that there
  exists a null geodesic $\gamma$ connecting $x$ to $y$ such that
  $k_{x}$ is coparallel and cotangent to $\gamma$ at $x$ and $k_{y}$
  is the parallel transport of $k_{x}$ from $x$ to $y$ along $\gamma$.
  The requirement $k_{x} \triangleright 0$ means that the covector
  $k_{x}$ is future directed.
\end{definition}

The above condition on the wave front set is rather useful and often
employed on practical grounds to check whether a given state really is
Hadamard or not.  Nonetheless, it is possible to provide another
definition via the so-called \textit{Hadamard form}, which has been
rigorously introduced in \cite{Kay:1991aa}.
\begin{definition}
  \label{Def:hadamardform}
  A state $\omega$ is said to be of the (local) \textbf{Hadamard form}
  if and only if in any convex normal neighbourhood the integral
  kernel of the associated two-point function can be written as
  \begin{equation*}
    \omega ( x , y ) = H ( x , y ) + W ( x , y ) \text{,}
  \end{equation*}
  where
  \begin{equation}
    \label{eq:singstr}
    H ( x , y ) = \lim_{\epsilon \to 0^{+}} \frac{U ( x , y
      )}{\sigma_{\epsilon} ( x , y )} + V ( x , y ) \ln
    \frac{\sigma_{\epsilon} ( x , y )}{\lambda^{2}} \text{,}
  \end{equation}
  and the limit is to be understood in the weak sense.  Here, $U$,
  $V$, as well as $W$ are smooth functions, while $\lambda$ is a
  reference length; furthermore,
  \begin{equation*}
    \sigma_{\epsilon} ( x , y ) \doteq \sigma ( x , y ) \pm 2i
    \epsilon \big( T ( x ) - T ( y ) \big) + \epsilon^{2}
  \end{equation*}
  with $\epsilon > 0$.  In the above formula, $T$ is a time function,
  such that $\nabla T$ is timelike and future directed on the full
  spacetime $( M , g )$.  In addition, if we apply \eqref{eq:eqm}
  either to the $x$- or to the $y$-variable, the result has to be a
  smooth function.
\end{definition}

The existence of a time function $T$ is guaranteed on any globally
hyperbolic manifold \cite{Bernal:2003aa,Bernal:2005aa} as these can be
decomposed as $\Sigma \times \mathbb{R}$, where $\Sigma$ is a smooth
Cauchy surface and $\mathbb{R}$ is the range of the time function $T$.

A completely satisfactory definition of the Hadamard form requires
some more work to rule out spacelike singularities, to circumvent
convergence problems of the series $V$, which is only asymptotic, and,
finally, to assure that the definition depends neither on a special
choice of the temporal function $T$ nor on the convex normal
neighbourhood employed.

In strict terms, we have only defined the local Hadamard form here.  A
stronger and more satisfactory definition, the so-called global
Hadamard form, has been introduced in \cite{Kay:1991aa}.  It
reinforces the local form extending it from the convex normal
neighbourhoods to certain ``causally-shaped'' neighbourhoods of a
Cauchy surface, thereby ruling out spacelike singularities.  However,
in \cite{Radzikowski:1996ab}, it has been shown that the local
Hadamard form already implies the global Hadamard form.

Another important fact is that the singular structure \eqref{eq:singstr}
is completely determined by the geometry of the background and the
equation of motion.  This of course does not hold for $W$ which encodes
the full state dependence.

\end{document}